\documentclass[aps,prd,onecolumn,nofootinbib,superscriptaddress]{revtex4-2}

\usepackage{amsmath}
\usepackage{amsfonts}
\usepackage{amssymb}
\usepackage{ulem}
\usepackage{color}
\usepackage[dvipsnames,svgnames]{xcolor}
\usepackage{graphicx}
\usepackage{dcolumn}
\usepackage{wrapfig,boxedminipage,setspace,subfigure}
\usepackage[title]{appendix}
\usepackage[colorlinks=true,linkcolor=blue,urlcolor=blue,filecolor=black,citecolor=red,pdfstartview=FitV,pdftitle={},pdfsubject={},pdfkeywords={},pdfpagemode=UseNone,bookmarksopen=true]{hyperref}
\usepackage[all]{hypcap}

\newcommand{\RNum}[1]{\uppercase\expandafter{\romannumeral #1\relax}}

\definecolor{ty}{RGB}{0, 160, 250}

\definecolor{yz}{RGB}{255, 0, 151}

\begin{document}
\baselineskip=0.5 cm

\title{Geodesic dynamics and multi-inclination images of a non-minimally coupled black~hole with a thin accretion disk}

\author{Tian-Yu Chen}
\email{chentianyu\_phys@foxmail.com}
\affiliation{Center for Gravitation and Cosmology, College of Physical Science and Technology, Yangzhou University, Yangzhou, 225009, China}

\author{Yong-Zhuang Li}
\email{liyongzhuang@just.edu.cn}
\affiliation{School of Science, Jiangsu University of Science and Technology, Zhenjiang, 212100, China}

\author{Xiao-Mei Kuang}
\email{xmeikuang@yzu.edu.cn}
\affiliation{Center for Gravitation and Cosmology, College of Physical Science and Technology, Yangzhou University, Yangzhou, 225009, China}

\begin{abstract}
\baselineskip=0.4 cm
In this paper, we investigate the optical properties of a black hole in non-minimal Einstein-Yang-Mills theory, illuminated by a thin accretion disk. In our setup, matter follows stable circular orbits outside the innermost stable circular orbit (ISCO), while inside the ISCO, it rapidly plunges into the black hole.  By analyzing the orbital dynamics of massive and massless particles, we find that the properties of both the ISCO and the photon sphere significantly depend on the non-minimal coupling parameter. Moreover, compared with the Schwarzschild and Reissner-Nordstr\"{o}m black holes, the non-minimal coupling extends the range of the impact parameter and slightly enhances the redshift effect in the images. Additionally, due to the significant influence of the non-minimal coupling parameter on the event horizon, the observed intensity of this black hole image under the selected emission model ultimately turns out to be weaker than that of the other two types of black holes, regardless of the inclination angle between the accretion disk and observation planes.
\end{abstract}

\maketitle
\tableofcontents
\newpage

\section{Introduction}\label{sec:introduction}

Recent groundbreaking observations, from the direct detection of gravitational waves emitted by coalescing black holes \cite{LIGOScientific:2016aoc,LIGOScientific:2016emj,LIGOScientific:2016lio}, to the resolved shadow images of M87* and Sgr A* observed by the Event Horizon Telescope (EHT) \cite{EventHorizonTelescope:2019dse,EventHorizonTelescope:2022wkp}, and the precise tracking of stellar dynamics in our Galactic Center \cite{GRAVITY:2019zy,GRAVITY:2021xju}, have collectively ushered in a new era of multimessenger astronomy. These observational techniques have not only provided direct and indirect evidence for the existence of black holes, but have also further demonstrated the success of Einstein's General Relativity (GR) in strong-field regimes. However, these findings do not rule out the potential validity of alternative theories of gravity, see \cite{Cai:2015emx,Bull:2015stt,Koyama:2015vza,Nojiri:2017ncd,Petrov:2020wgy,CANTATA:2021asi,Odintsov:2022cbm,Shankaranarayanan:2022wbx,Bambi:2023jiz,Yunes:2024lzm} for a comprehensive introduction to the modified theories of gravity. Among these, the Einstein-Yang-Mills (EYM) theory, which combines Einstein's theory of gravity with non-Abelian gauge theory, has consistently been a subject of considerable research attention.

One of the natural and fundamentally significant motivations for studying EYM theory lies in the fact that neither the vacuum Einstein equations nor pure Yang-Mills theory in 3+1 dimensions admit nontrivial, static, globally regular solutions \cite{Einstein:1943ixi,Deser:1976wq,Coleman:1977hd}. In contrast, such monopole solutions do emerge within the coupled EYM framework, see \cite{Bartnik:1988am,Lavrelashvili:1992ia,Smoller:1993bs,Smoller:1995nk,Kleihaus:1996vk,Kleihaus:1997rb,Hosotani:1999yg,Bjoraker:1999yd,Hosotani:2000fu,Bjoraker:2000qd,Kleihaus:2003sh,Breitenlohner:2005hx,Brihaye:2006xc,Mazharimousavi:2008ap,HabibMazharimousavi:2008ib,Devecioglu:2014iia,Edery:2018jyp,Altas:2021htf} and references therein. On the one hand, the spacetime curvature of these regular black hole solutions remains finite everywhere, thereby avoiding the curvature singularity problem inherent in classical Einstein gravity \cite{Lan:2023cvz}. On the other hand, the non-Abelian gauge field used to describe matter outside the regular event horizon manifests as black hole hair \cite{Kleihaus:1998sm}, which can also be regarded as counterexamples to the black hole no-hair theorem. Building upon this foundation, the non-minimal coupling between the gravitational and gauge fields subsequently provides physicists with an expanded set of physical phenomena for theoretical and observational investigation \cite{Hehl:1999bt,Balakin:2010ar,Bergliaffa:2020ivp}, and most importantly some exact regular black hole solutions have been constructed in the non-minimally coupled EYM theory with a $SU(2)$ Wu-Yang ansatz \cite{Mueller-Hoissen:1987nvb,Balakin:2006gv,Balakin:2015oea,Balakin:2015gpq}.

Early-stage investigations have been conducted into the fundamental characteristics of regular EYM black holes, especially on the black hole images, particle motions, and quasinormal modes (QNMs). Following the regular non-minimal magnetic black hole solution in a spacetime with cosmological constant formulated in Ref.~\cite{Balakin:2015gpq}, Azam et al. systematically investigated the geodesic structure for both massive particles and massless photons in this gravitational background \cite{Azam:2017izk}. Such particle dynamics were re-examined by Al-Badawi and Owaidat in Ref.~\cite{Al-Badawi:2023jtq} without the cosmological constant, where they also provided the horizon structure, photon sphere radius, and inner stable circular orbit (ISCO) of a mass particle with EYM parameters. Liu et al. investigated the weak and strong deflection gravitational lensings by the regular non-minimal EYM black hole \cite{Liu:2019pov}, and similar analyses can also be found in Ref.~\cite{Bergliaffa:2020ivp} where the authors examined the lensing at large deflection angles caused by a Schwarzschild black hole for the case of a non-minimal coupling between gravitation and electromagnetism. Jusufi et al. obtained an effective metric describing a rotating regular magnetic black hole in non-minimally coupled EYM theory using the Newman-Janis algorithm via the non-complexification radial coordinate procedure, and studied its shadow, QNMs and quasiperiodic oscillations (QPOs) in Ref.~\cite{Jusufi:2020odz}. The shadow and weak gravitational lensing for such a rotating regular black hole in the presence of plasma have been investigated in Ref.~\cite{Kala:2022uog}, while the strong gravitational lensing was investigated in Ref.~\cite{,Zhang:2023oui}, suggesting that the influence of magnetic charge and EYM parameters is small. Very recently, the scalar QNMs using the higher-order WKB method and black hole shadow for the regular solution of Ref.~\cite{Balakin:2015gpq} was presented in Ref.~\cite{Gogoi:2024vcx}. However, it was demonstrated that this analysis lacks sufficient accuracy when studying modes with $l\leq n$, where $l$ is the multipole number and $n$ is the overtone number \cite{Lutfuoglu:2025ljm}. More investigations on the QNMs, Hawking radiation, or grey-body factors of the EYM black hole \cite{Balakin:2015gpq} have been presented in Refs.~\cite{Al-Badawi:2023emj,Pu:2023hll,Ji:2025nlc,Dubinsky:2025bvf}.

The optical appearances of black holes provide a valuable window into exploring their background spacetime \cite{Luminet:1979nyg,Falcke:1999pj,Tsukamoto:2014tja,Cunha:2015yba,Tsukamoto:2017fxq,Cunha:2018acu,Perlick:2021aok,Chen:2022scf}. However, the imaging features of the black hole are influenced not only by the background spacetime geometry but also significantly by the accretion material surrounding it, particularly the luminous accretion flow \cite{Gralla:2019xty}. As expected, to accurately describing the characteristics of the black hole's image requires sophisticated numerical computations, specifically, general relativistic magnetohydrodynamic (GRMHD) simulations, due to the complex physical environment near the event horizon of a black hole \cite{EventHorizonTelescope:2019pcy,Pugliese:2020syx}. However, simplified accretion models are often sufficient for capturing the primary features of a black hole image \cite{Abramowicz:2011xu} and numerous studies have investigated the image characteristics of black holes with simplified accretion disks within various theoretical frameworks, see \cite{Dokuchaev:2019pcx,Narayan:2019imo,Zeng:2020dco,Saurabh:2020zqg,Qin:2020xzu,Eichhorn:2021iwq,He:2021htq,Li:2021riw,Liu:2021yev,Hou:2022eev,Wang:2023vcv,Meng:2023htc,Zeng:2023fqy,Hu:2023pyd,Wang:2023fge,DeMartino:2023ovj,Sui:2023tje,Gao:2023mjb,Heydari-Fard:2023kgf,Meng:2024puu,Donmez:2024lfi,Zare:2024dtf,Wang:2024lte,Guo:2024mij,Yang:2024nin,Meng:2025ivb,Lim:2025cne} and references therein. Our aim here is to investigate the optical appearances of the regular black hole illuminated by a thin accretion disk in the non-minimally coupled EYM theory, as relevant studies remain notably scarce to date. 
To this end, we shall construct a dynamical thin accretion disk extending to the event horizon of the black hole as the light source. Specifically, we consider a simple, thin, Keplerian accretion disk composed of free, electrically neutral plasma, where the motion of the particle follows circular orbits outside the inner stable circular orbit (ISCO), while inside the ISCO, the accretion flow can quickly plunge into the black hole.

This paper is organized as follows: in Sec.~\ref{sec:black_hole}, we provide a concise review of the non-minimally coupled EYM theory with an exact spherically symmetric black hole solution derived therefrom. Then in Sec.~\ref{sec:particle_motion} we examine the orbital motion of massive particles and photons in this black hole spacetime, laying the foundation for investigating the image features of a black hole surrounded by a thin accretion disk in Sec.~\ref{sec:disk_images}. Finally, we conclude with a summary of the results and a discussion of future prospects in Sec.~\ref{sec:conclusion}. In this paper, we will adopt the natural unit with $G = c = 1$ unless we restate.

\section{Brief review of non-minimal Einstein-Yang-Mills theory}\label{sec:black_hole}

The action of non-minimal EYM theory with $SU(2)$ symmetry and Wu-Yang-type ansatz is described as follows \cite{Balakin:2015gpq},
\begin{equation}
\label{eq:action}
	S_{\mathrm{NMEYM}} = \int \mathrm{d}^{4}x \sqrt{-g} \left( \frac{R}{8\pi} + \frac{1}{2} F^{(i)}_{\mu \nu} F^{\mu \nu (i)} + \frac{1}{2} \mathfrak{R}^{\alpha \beta \mu \nu} F^{(i)}_{\alpha \beta} F^{(i)}_{\mu \nu} \right),
\end{equation}
where the Greek indices range from $0$ to $3$, and the Latin (group) indices range from $1$ to $3$. $R$ is the Ricci scalar. Note that repeated group indices imply summation with a Kronecker delta tensor. The $SU(2)$ Yang-Mills field is represented by a triplet of vector potentials $A^{(i)}_\mu$ which are related to the Yang-Mills field components $F^{(i)}_{\mu \nu}$ through the formula given below,
\begin{equation}
\label{eq:vector_potential}
	F^{(i)}_{\mu \nu} = \nabla_{\mu} A^{(i)}_{\nu} - \nabla_{\nu} A^{(i)}_{\mu} + f^{(i)}_{~(j)(k)} A^{(j)}_{\mu} A^{(k)}_{\nu}.
\end{equation}
Here $\nabla_{\mu}$ is the covariant derivative, and the symbols $f^{(i)}_{~(j)(k)}$ represent the real structure constants of the $3$-parameter Yang-Mills gauge group $SU(2)$. The non-minimal susceptibility tensor $\mathfrak{R}^{\alpha \beta \mu \nu}$ is defined as
\begin{equation}
\label{eq:tensor}
\mathfrak{R}^{\alpha \beta \mu \nu} = \frac{q_{1}}{2} R \left( g^{\alpha \mu} g^{\beta \nu} - g^{\alpha \nu} g^{\beta \mu} \right) + \frac{q_{2}}{2} \left( R^{\alpha \mu} g^{\beta \nu} - R^{\alpha \nu} g^{\beta \mu} + R^{\beta \nu} g^{\alpha \mu} - R^{\beta \mu} g^{\alpha \nu} \right) + q_{3} R^{\alpha \beta \mu \nu},
\end{equation}
where $R^{\mu \nu}$, $R^{\alpha \beta \mu \nu}$, and $q_{1}, q_{2}, q_{3}$ represent the Ricci tensor, Riemann tensor, and phenomenological parameters that describe the non-minimal coupling between the Yang-Mills field and the gravitational field. The above action \eqref{eq:action} admits a static spherically symmetric spacetime described by the metric \cite{Balakin:2015gpq}
\begin{align}
\label{eq:metric-original}
	\mathrm{d}s^2 = -f(r) \mathrm{d}t^2 + \frac{\mathrm{d}r^2}{f(r)} + r^2 ( \mathrm{d}\theta^2 + \sin^2\theta \mathrm{d}\phi^2 ) \nonumber \\
	\text{with} \quad f(r) = 1 + \left( \frac{r^4}{r^4 + 2q Q^2} \right) \left( \frac{Q^2}{r^2} - \frac{2M}{r} \right),
\end{align}
where $q_1, q_2, q_3$ are related to a sole independent parameter $q$ introduced by the spherical symmetry of the solution. Clearly, when $q$ approaches zero, the metric reduces to that of the Reissner-Nordstr\"{o}m (RN) black hole. We will only consider $q>0$ here. By scaling the coupling constant $q$ with $Q^2$, the term $q Q^2$ in the metric function \eqref{eq:metric-original} can be replaced by $\xi$. Thus, the metric can be rewritten as
\begin{align}
\label{eq:metric}
	\mathrm{d}s^2 = -f(r) \mathrm{d}t^2 + \frac{\mathrm{d}r^2}{f(r)} + r^2 ( \mathrm{d}\theta^2 + \sin^2\theta \mathrm{d}\phi^2 ) \nonumber \\
	\text{with} \quad f(r) = 1 + \left( \frac{r^4}{r^4 + 2\xi} \right) \left( \frac{Q^2}{r^2} - \frac{2M}{r} \right).
\end{align}
Then, the parameters $M$ and $Q$ are related with the mass and charge of the black hole, while $\xi$ describes the coupling strength between the Yang-Mills field and the gravitational field. It is noted that we shall consider the dimensionless black hole parameters $Q/M$ and $\xi/M^4$. All physical quantities in our study will be rescaled to be dimensionless by $M$ and we set $M = 1$ in the calculations.

\begin{figure}[htbp]
	\centering
	\includegraphics[height=6.5cm]{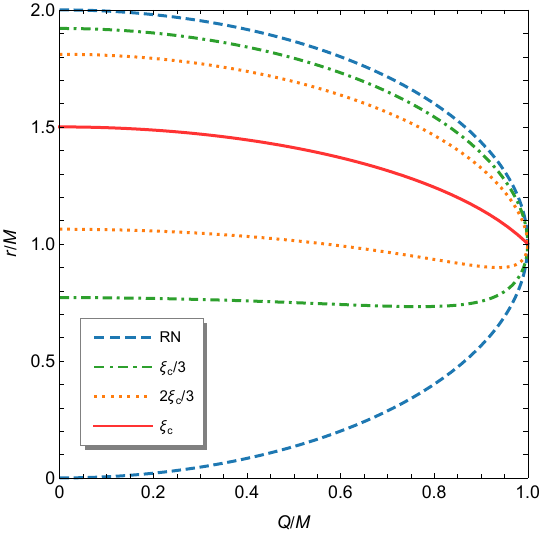}
	\caption{The radii of both the Cauchy and event horizons as functions of $Q$, for $\xi = 0~\text{(RN)}$, $\xi = \frac{1}{3} \xi_{\mathrm{c}}$, $\xi = \frac{2}{3} \xi_{\mathrm{c}}$, and $\xi = \xi_{\mathrm{c}}$, which respectively correspond to the blue dashed, green dot-dashed, orange dotted, and red solid lines.}
    \label{fig:horizons}
\end{figure}

To ensure the existence of horizons, we note that a critical relationship must hold between $\xi$ and $Q$. Solving $f(r) = f'(r) = 0$ simultaneously, one has
\begin{align}
\label{eq:critical_xi_Q}
    &r_{\mathrm{c}} = \frac{1}{4} \left( 3M + \sqrt{9M^2 - 8Q^2} \right), \\
    &\xi_{\mathrm{c}} = \frac{1}{64} \left( 27M^{4} - 36M^2 Q^2 + 8Q^4 + M \sqrt{(9M^2 - 8Q^2)^{3}} \right).
\end{align}
Within the framework of our model, the assumption $q>0$ necessarily implies $\xi>0$. Consequently, in determining the critical value, all solutions yielding $\xi\leq 0$ have been excluded. This leads to the conclusion that the metric \eqref{eq:metric} admits exactly one horizon $r_{\mathrm{c}}$ under critical conditions. Furthermore, for arbitrary values of $Q$ and $M$ satisfying $2\sqrt{2}Q \leq 3M $,\footnote{Noted that we further require $Q < M$ to guarantee $\xi_{\mathrm{c}} > 0$.} the existence of the horizons requires $\xi \leq \xi_{\mathrm{c}}$, establishing an upper bound for the parameter in the black hole solution, see Fig.~\ref{fig:horizons}, where the Cauchy horizon and the event horizon will coincide at $\xi = \xi_{\mathrm{c}}$. Meanwhile, as previously mentioned, under the assumption of $q>0$, the black hole spacetime described by the metric \eqref{eq:metric} is singularity-free. Specifically, no radial coordinate value exists that would cause divergence of the Kretschmann scalar $\mathcal{K} = R^{\alpha \beta \mu \nu} R_{\alpha \beta \mu \nu}$.

\section{Particle motion for the non-minimally coupled black hole}\label{sec:particle_motion}

In this section, we provide a concise description of the geodesic characteristics for particles orbiting the non-minimally coupled black hole \eqref{eq:metric}, encompassing both timelike geodesics of massive particles and null geodesics of photons. The former serves as a fundamental element in our construction of the accretion disk, while the latter provides essential information required for characterizing the black hole's imaging properties.

\subsection{General formula for particle motion}\label{subsec:general_formula}

Here, we adopt the description from Chandrasekhar \cite{Chandrasekhar:1985kt}, according to which the Lagrangian for particle motion in a general spherically symmetric spacetime can be written as
\begin{equation}
	\mathcal{L} = \frac{1}{2} g_{\mu \nu} \dot{x}^{\mu} \dot{x}^{\nu},
\end{equation}
and the corresponding Hamiltonian is
\begin{equation}
\label{eq:Hamitolian}
	\mathcal{H} = p_{\mu} \dot{x}^{\mu} - \mathcal{L},
\end{equation}
where $p_\mu$ is the conjugate momentum of the particles, defined by
\begin{equation}
\label{eq:p_mu}
	p_{\mu} = g_{\mu\nu} \dot{x}^{\nu}.
\end{equation}
The dot over a symbol denotes the derivative with respect to an affine parameter $\lambda$. By relating the proper time $\tau = \delta\lambda$ one has the normalizing condition
\begin{equation}
\label{eq:Lagrangian}
    \mathcal{L} = \frac{1}{2} g_{\mu \nu} \dot{x}^{\mu} \dot{x}^{\nu} = -\frac{1}{2} \delta^2,
\end{equation}
where $\delta$ is the particle rest mass, with $\delta^{2} = 0\ \text{and}\ 1$ giving the null and timelike geodesics, respectively. Due to time-translational and rotational invariance of the black hole, two conserved quantities appear in this dynamic system, which correspond to the specific energy $E$ and angular momentum $L$ of the particle,
\begin{equation}
\label{eq:E-L-t}
	E = -g_{tt} \dot{t}, \quad L = g_{\phi \phi} \dot{\phi}.
\end{equation}
Considering the spherical symmetry, we can conveniently set $\theta = \pi/2,\ \dot{\theta} = 0$. Thus, according to Eqs.~\eqref{eq:metric}, \eqref{eq:Lagrangian}, and \eqref{eq:E-L-t}, one can obtain the radial equation of particle motion as
\begin{equation}
\label{eq:radial_eq}
	\dot{r}^2 = E^2 - f(r) \left( \delta^{2} + \frac{L^2}{r^2} \right),
\end{equation}
from which we can define the effective potential as
\begin{equation}
\label{eq:effective_potential}
	V_{\mathrm{eff}} = E^2 - \dot{r}^2 = f(r) \left( \delta^{2} + \frac{L^2}{r^2} \right).
\end{equation}
The following analyses of the particles' motion are based on the above general setup.

\subsection{Motion of massive particles and the innermost stable circular orbit}\label{subsec:massive_particle}

\begin{figure}[htbp]
	\centering
	{\includegraphics[width=5.5cm]{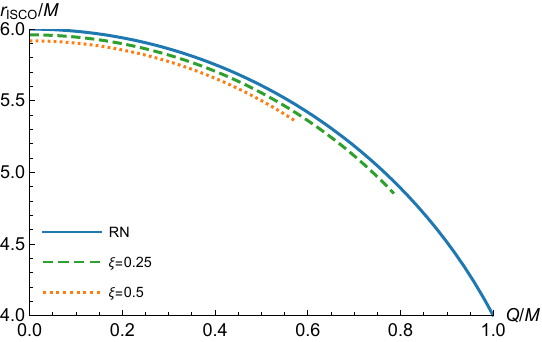}}\hspace{0.5cm}
	{\includegraphics[width=5.5cm]{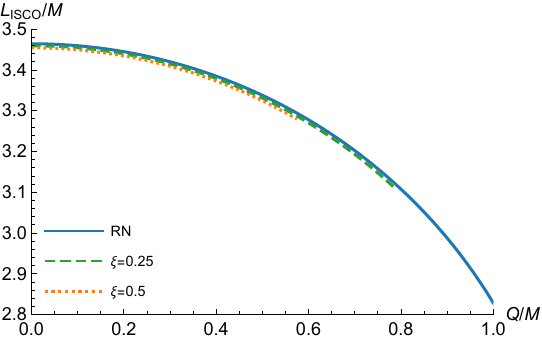}}\hspace{0.5cm}
	{\includegraphics[width=5.5cm]{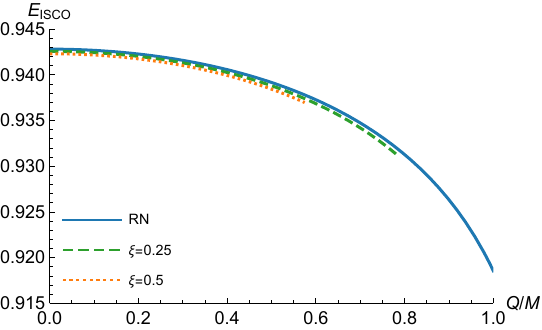}} \\
    \subfigure[\, $r_{\mathrm{ISCO}}$]
	{\includegraphics[width=5.5cm]{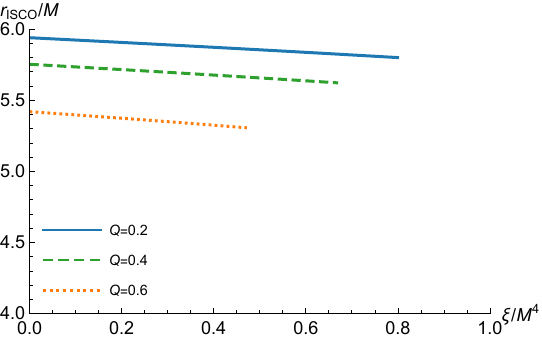}}\hspace{0.5cm}
    \subfigure[\, $L_{\mathrm{ISCO}}$]
	{\includegraphics[width=5.5cm]{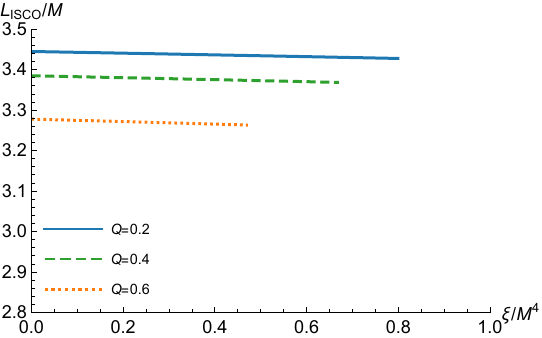}}\hspace{0.5cm}
    \subfigure[\, $E_{\mathrm{ISCO}}$]
	{\includegraphics[width=5.5cm]{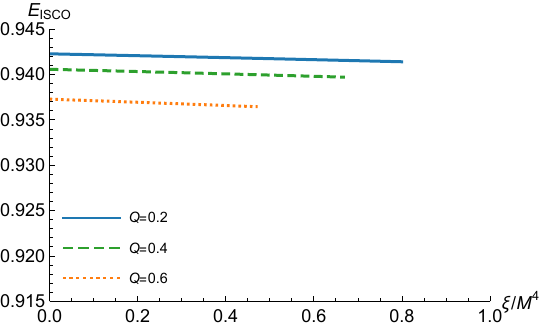}}
	\caption{The radius, angular momentum, and energy of a particle in the ISCO, as functions of $Q$ with $\xi$ fixed (top row), and as functions of $\xi$ with $Q$ fixed (bottom row).}
    \label{fig:ISCO}
\end{figure}

We now consider the massive particles where $\delta^2 = 1$. Subsequently, the ISCO is determined by the conditions,
\begin{equation}
	V_{\mathrm{eff}}(r) = E^2, \quad \frac{\mathrm{d}V_{\mathrm{eff}}}{\mathrm{d}r} = 0, \quad \frac{\mathrm{d}^{2}V_{\mathrm{eff}}}{\mathrm{d}r^2} = 0,
\end{equation}
from which we can determine the radius $r_{\mathrm{ISCO}}$, the angular momentum $L_{\mathrm{ISCO}}$, and the energy $E_{\mathrm{ISCO}}$ of the particle as functions of the black hole parameters, see Fig.~\ref{fig:ISCO}. As shown in the figure, the values of $r_{\mathrm{ISCO}}$, $L_{\mathrm{ISCO}}$, and $E_{\mathrm{ISCO}}$ all decrease as the black hole parameters $Q$ or $\xi$ increase. In the limit where both $Q$ and $\xi$ tend to zero, the values of all three quantities reduce to their counterparts for a Schwarzschild black hole, i.e., $(r_{\mathrm{ISCO}}, L_{\mathrm{ISCO}}, E_{\mathrm{ISCO}}) = (6, 2\sqrt{3}, 2\sqrt{2}/3)$. Such behaviors admit a clear physical interpretation. Indeed, the fact $r_{\mathrm{ISCO}}^4 \gg 1$ indicates that for small values of $\xi$, the coefficient $r^4/(r^4 + 2\xi)$ in the metric function $f(r)$ remains close to one. As a result, the influence of $\xi$ is not significant, and the dependence of $r_{\mathrm{ISCO}}$, $L_{\mathrm{ISCO}}$, and $E_{\mathrm{ISCO}}$ on $Q$ should closely resemble the behavior described in Ref.~\cite{Pugliese:2010ps} for the classical RN black hole. In particular, with $Q$ held fixed, all three quantities decrease approximately linearly with increasing $\xi$. 

\begin{figure}[htbp]
	\centering
    \subfigure[\, $E$ of massive particles in SCOs]
	{\includegraphics[width=5.5cm]{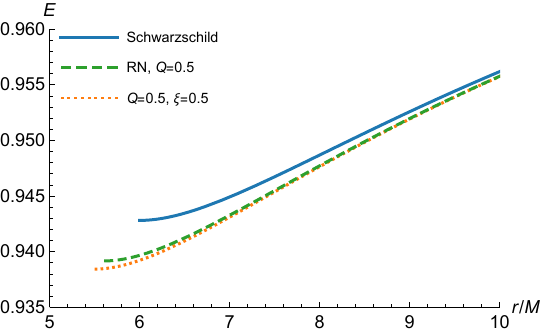}}\hspace{0.5cm}
    \subfigure[\, $L$ of massive particles in SCOs]
	{\includegraphics[width=5.5cm]{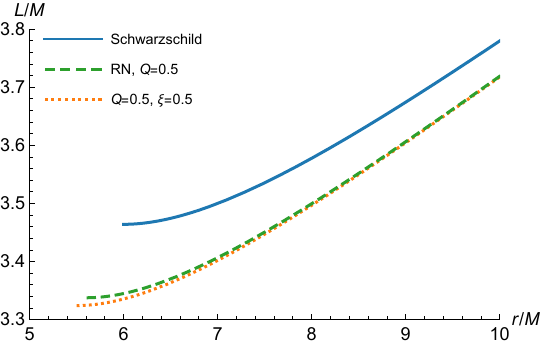}}\hspace{0.5cm}
    \subfigure[\, $\Omega$ of massive particles in SCOs]
	{\includegraphics[width=5.5cm]{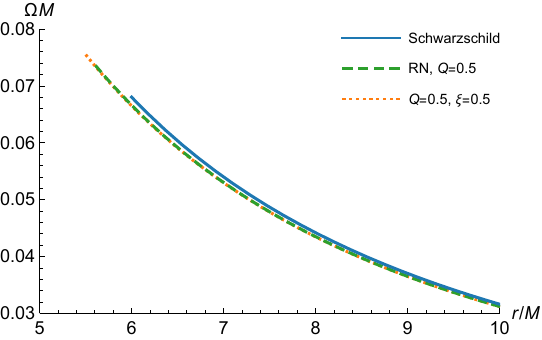}}\hspace{0.5cm}
	\caption{The profiles of the energy, angular momentum, and angular velocity of particles in stable circular orbits for the Schwarzschild, RN, and non-minimally coupled black holes.}
    \label{fig:SCO}
\end{figure}

Furthermore, to gain deeper insight into the properties of circular orbits in the black hole spacetime under investigation, we present a detailed analysis of how the stable circular orbits (SCOs) are influenced by $Q$ or $\xi$. Such SCOs require $V_{\mathrm{eff}}''(r) > 0$. By introducing the orbital angular velocity $\Omega = \mathrm{d}\phi / \mathrm{d}t = \dot{\phi} / \dot{t}$, the final expressions for the energy, angular momentum, and angular velocity of a particle in the SCO are obtained as follows:
\begin{align}
\label{eq:SCO_E-L-O}
    &E = \frac{f(r)}{\sqrt{f(r) - r^2 \Omega^2}}, \nonumber \\
    &L = \frac{r^2 \Omega}{\sqrt{f(r) - r^2 \Omega^2}}, \\
    &\Omega = \frac{\sqrt{Mr \left( r^4 - 6\xi \right) - Q^2 \left( r^4 - 2\xi \right)}}{r^4 + 2\xi}. \nonumber
\end{align}
The profiles of the energy, angular momentum, and angular velocity of particles in stable circular orbits for the Schwarzschild, RN, and non-minimally coupled black holes are shown in Fig.~\ref{fig:SCO}, which shows that they are marginally affected by $\xi$ with the selected values in the current investigation.

\subsection{Null geodesic and photon trajectories}\label{subsec:photon_trajectory}

Light rays emitted by the accretion disk will be bent by the central black hole around which they orbit. These photons are either absorbed by the black hole or scattered to infinity, with only a minute fraction ultimately reaching a distant observer. These light rays received by the observer will form the black hole images with different orders, and can be simulated by tracing the light rays backward in time from the observer's image screen to the light source near the black hole. Here, we use the ray tracing algorithms addressed in Refs.~\cite{Gralla:2020srx,Hou:2022eev}.

\begin{figure}[htbp]
	\centering
	\includegraphics[width=6.5cm]{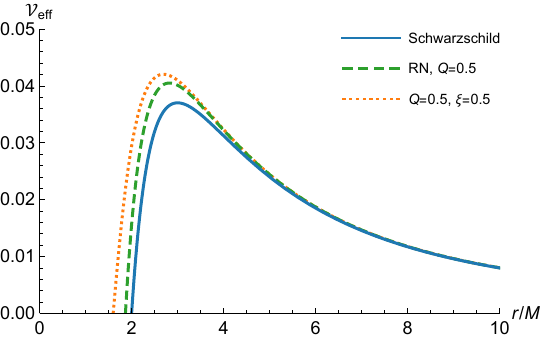}
	\caption{The effective potential of photons for the Schwarzschild, RN, and non-minimally coupled black holes.}
    \label{fig:Veff}
\end{figure}

Based on Eqs.~\eqref{eq:E-L-t}, \eqref{eq:radial_eq}, \eqref{eq:effective_potential}, and the definition of the impact parameter $b \equiv L/E$, we can obtain the equations of motion for photons around the black hole:
\begin{align}
	\dot{t} &= \frac{1}{b f(r)},\label{eq:t} \\
	\dot{\phi} &= \pm \frac{1}{r^2},\label{eq:phi} \\
	\dot{r}^2 &= \frac{1}{b^2} - \mathcal{V}_{\mathrm{eff}}(r),\label{eq:r}
\end{align}
where the sign $\pm$ in Eq.~\eqref{eq:phi} depends on the direction of motion, i.e., clockwise or counterclockwise, and the effective potential of the photon is now
\begin{equation}
\label{eq:Veff}
	\mathcal{V}_{\mathrm{eff}}(r) = \frac{f(r)}{r^2}.
\end{equation}
Note that here we have redefined the affine parameter $\lambda$ as $\tilde{\lambda} = \lambda L$ then the dots in Eqs.~\eqref{eq:t}--\eqref{eq:r} are with respect to $\tilde{\lambda}$ now \cite{Wang:2023vcv}. Apparently, with fixed $Q$, the larger $\xi$ leads to larger $\mathcal{V}_{\mathrm{eff}}(r)$, as shown in Fig.~\ref{fig:Veff}. 

\begin{figure}[htbp]
	\centering
	{\includegraphics[width=5.5cm]{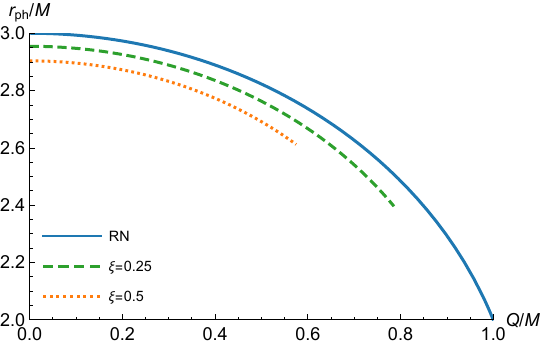}}\hspace{0.5cm}
	{\includegraphics[width=5.5cm]{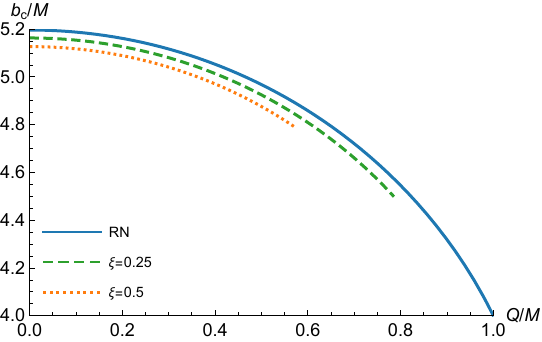}}\hspace{0.5cm}
	{\includegraphics[width=5.5cm]{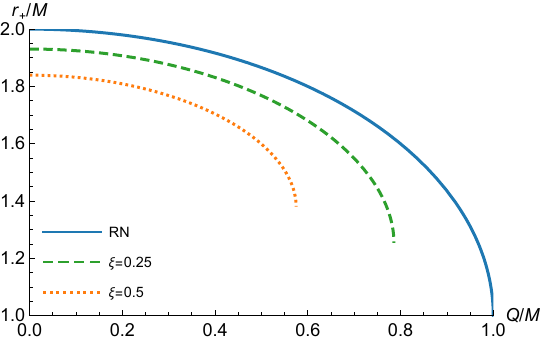}} \\
    \subfigure[\, $r_{\mathrm{ph}}$]
	{\includegraphics[width=5.5cm]{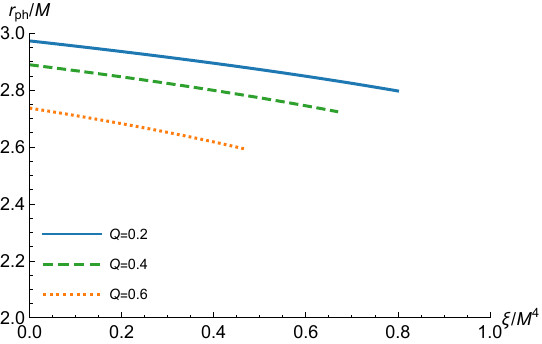}}\hspace{0.5cm}
    \subfigure[\, $b_{\mathrm{c}}$]
	{\includegraphics[width=5.5cm]{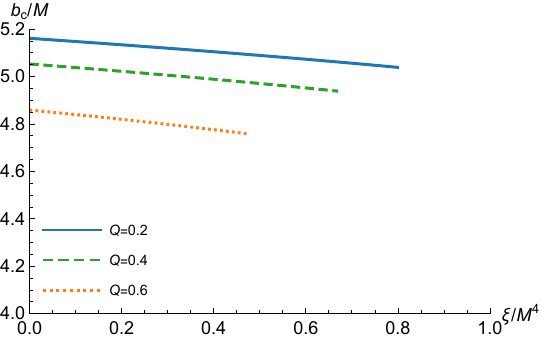}}\hspace{0.5cm}
    \subfigure[\, $r_{+}$]
	{\includegraphics[width=5.5cm]{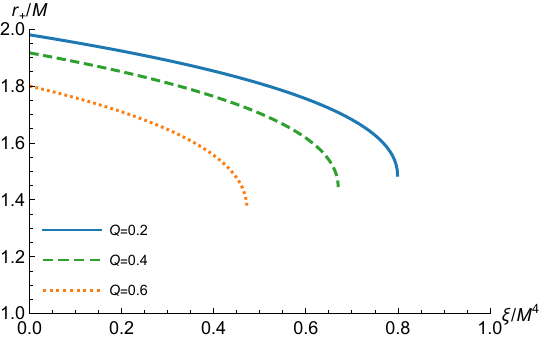}}
	\caption{The radius of the photon sphere, the critical impact parameter, and the radius of the event horizon, as functions of $Q$ with $\xi$ fixed (top row), and as functions of $\xi$ with $Q$ fixed (bottom row).}
    \label{fig:rph+bc+reh}
\end{figure}

The radial geodesic equation \eqref{eq:r} determines the fate of a photon based on its impact parameter. In particular, the critical impact parameter $b_{\mathrm{c}}$ can be determined according to Eq.~\eqref{eq:r} as
\begin{equation}
\label{eq:rph-bc}
	\mathcal{V}_{\mathrm{eff}}(r_\mathrm{ph}) = \frac{1}{{b_{\mathrm{c}}}^2}, \quad \mathcal{V}_{\mathrm{eff}}'(r_\mathrm{ph}) = 0.
\end{equation}
The photon orbits with $b = b_{\mathrm{c}}$ correspond to the photon sphere at radius $r_{\mathrm{ph}}$. These orbits are unstable, meaning that a small radial perturbation will either cause the photon to escape to infinity ($b > b_{\mathrm{c}}$) or be captured by the black hole ($b < b_{\mathrm{c}}$). We then numerically solve Eq.~\eqref{eq:rph-bc} to determine the photon sphere and the critical impact parameter as functions of the model parameters, see Fig.~\ref{fig:rph+bc+reh}. As shown in the figure, the radius $r_{\mathrm{ph}}$ of the photon sphere and the corresponding critical impact parameter $b_c$ are influenced by $Q$ and $\xi$ in a manner analogous to how these parameters affect massive particles in the ISCO in Fig.~\ref{fig:ISCO}, though the effect on photons is more pronounced. In contrast, the dependence of the event horizon radius $r_{+}$ on $\xi$ deviates from an approximately linear relationship.

\begin{figure}[htbp]
	\centering
    \includegraphics[height=6.5cm]{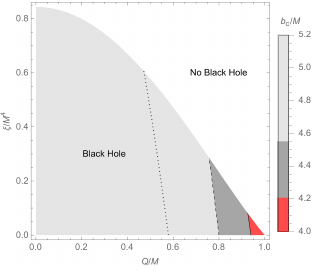}
	\caption{The constraint on the black hole parameters $Q$ and $\xi$. The colored region denotes the parameter regime satisfying the condition for the existence of black holes \eqref{eq:mc}, while no black hole exists in the white region. The lighter gray region represents the $1\sigma$ constraint \eqref{eq:1sigma_constraint}, while the darker gray region corresponds to the $2\sigma$ constraint \eqref{eq:2sigma_constraint}. The solid, dashed, and dotted lines respectively correspond to $b_{\mathrm{c}} = 4.21,\ 4.55,\ \text{and}\ 4.88$.}
    \label{fig:constraint}
\end{figure}

In Sec.~\ref{sec:black_hole} we have stated that the black hole parameters $(\xi, Q, M)$ have the relation \eqref{eq:critical_xi_Q} to ensure the existence of the horizon. Such relation can be rewritten as \cite{Balakin:2015gpq},
\begin{equation}
\label{eq:mc}
	M_{\mathrm{c}} = \sqrt{2 Q^2} \frac{\left( 1 + \frac{8\xi}{Q^4} + \sqrt{1 + \frac{24\xi}{Q^4}} \right)}{\left( 1 + \sqrt{1 + \frac{24\xi}{Q^4}} \right)^{\frac{3}{2}}} \le M,
\end{equation}
where $M_{\mathrm{c}}$ is defined as the critical mass for given $(\xi, Q)$. This relationship can be visualized more clearly and intuitively through the diagram presented in Fig.~\ref{fig:constraint}. In terms of observation, these two parameters $Q$ and $\xi$ can be further constrained from the black hole shadow radius, which is equal to the critical impact parameter $b_\mathrm{c}$. According to the observations of Sgr A* by the EHT, we have the following constraints on $b_\mathrm{c}/M$ \cite{Vagnozzi:2022moj},
\begin{align}
	4.55 \lesssim b_\mathrm{c}/M \lesssim 5.22 \quad (1\sigma),\label{eq:1sigma_constraint} \\
	4.21 \lesssim b_\mathrm{c}/M \lesssim 5.56 \quad (2\sigma).\label{eq:2sigma_constraint}
\end{align}
As shown in Fig.~\ref{fig:constraint}, the lighter gray region represents the $1\sigma$ constraints \eqref{eq:1sigma_constraint}, while the darker gray region corresponds to the $2\sigma$ constraints \eqref{eq:2sigma_constraint}. The solid, dashed, and dotted lines correspond to $b_{\mathrm{c}} = 4.21,\ 4.55,\ \text{and}\ 4.88$, respectively. We can see that the large charge solutions are ruled out by the constraint, similar to the finding in Ref.~\cite{EventHorizonTelescope:2021dqv}. The choice of black hole parameters in the current study focuses on the $1\sigma$ constraint regime.

\begin{figure}[htbp]
	\centering
    \subfigure[\, $\text{Schwarzschild}$]
	{\includegraphics[width=5.5cm]{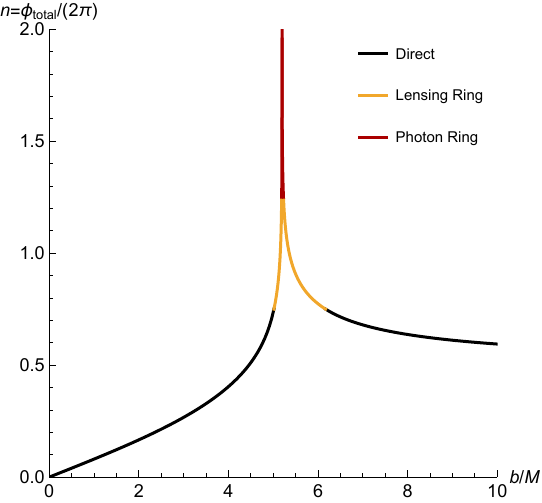}}\hspace{0.5cm}
    \subfigure[\, $\text{RN}, Q = 0.5$]
	{\includegraphics[width=5.5cm]{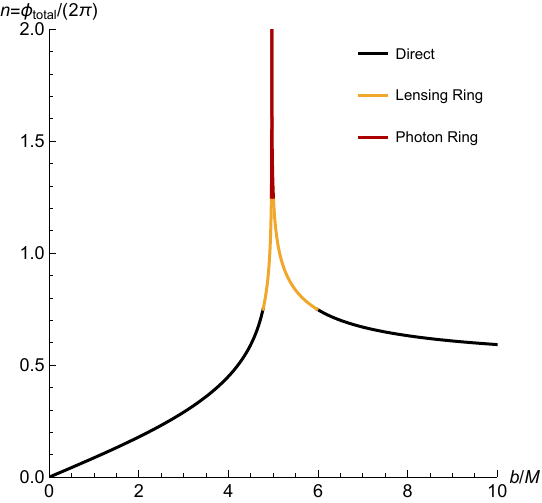}}\hspace{0.5cm}
    \subfigure[\, $Q = 0.5, \xi = 0.5$]
    {\includegraphics[width=5.5cm]{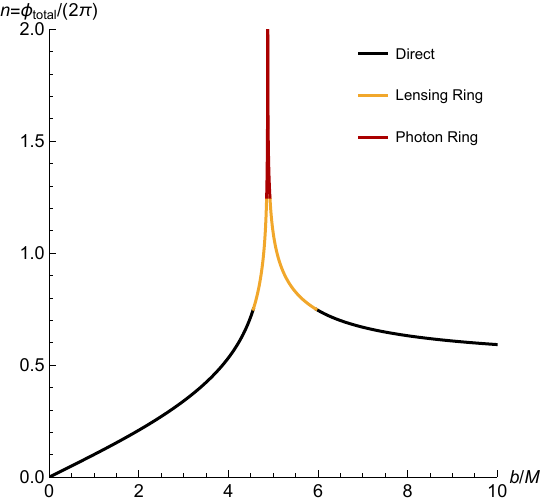}} \\
    \subfigure[\, $\text{Schwarzschild}$]
	{\includegraphics[width=5.5cm]{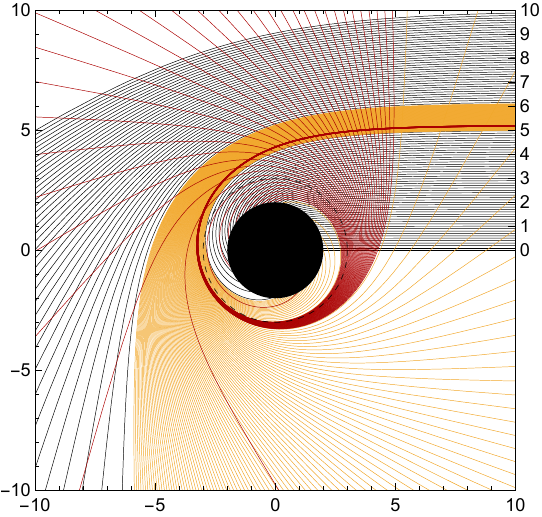}}\hspace{0.5cm}
    \subfigure[\, $\text{RN}, Q = 0.5$]
	{\includegraphics[width=5.5cm]{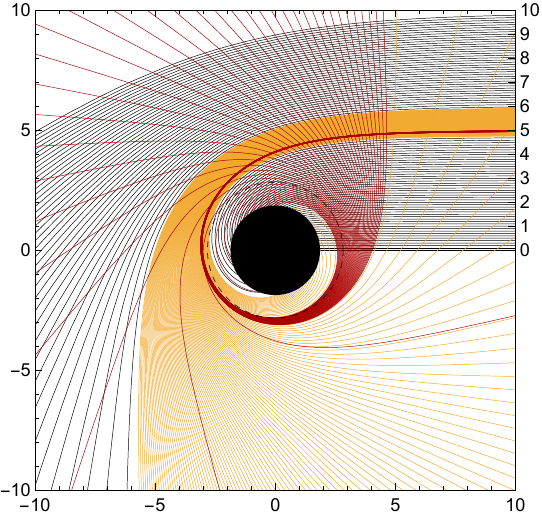}}\hspace{0.5cm}
    \subfigure[\, $Q = 0.5, \xi = 0.5$]
    {\includegraphics[width=5.5cm]{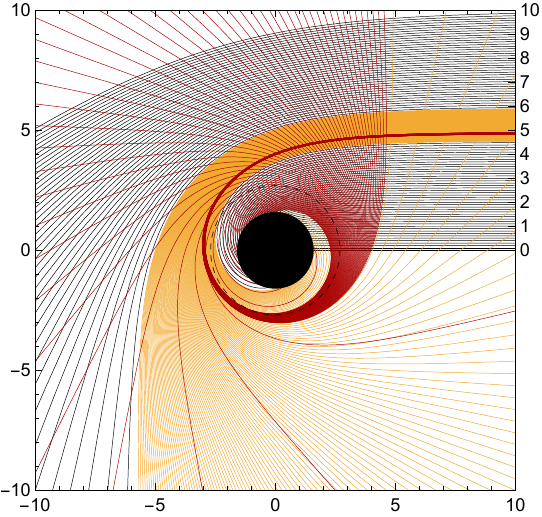}}
	\caption{\textbf{Top row:} the orbit numbers $n$ as functions of the impact parameter $b$ for the Schwarzschild, RN, and non-minimally coupled black holes. \textbf{Bottom row:} the distribution of light rays around the the Schwarzschild, RN, and non-minimally coupled black holes. The solid black disk stands for the black hole and the black dashed circle denotes the photon sphere, while the black, gold, and red curves correspond to the direct emission, lensing ring emission, and photon ring emission, respectively.}
    \label{fig:nb+trajectories}
\end{figure}

We now move on to study the trajectory of photon, which can be depicted by the orbit equation obtained from Eqs.~\eqref{eq:phi} and \eqref{eq:r} as
\begin{equation}
\label{eq:r-phi}
	\frac{\mathrm{d}r}{\mathrm{d}\phi} = \pm r^2 \sqrt{\frac{1}{b^2} - \mathcal{V}_{\mathrm{eff}}(r)}.
\end{equation}
Clearly, the trajectory of a photon is determined by the effective potential $\mathcal{V}_{\mathrm{eff}}$ and the impact parameter $b$. Then we can employ the backward ray-tracing method to visualize the trajectories. In this approach, the trajectory of each photon arriving at the observer's screen is traced backward to determine its point of origin. Depending on the impact parameter $b$, photons may complete different numbers of orbits around the central black hole before reaching the observer. Subsequently, the impact parameter region can be classified according to the total number of photon orbits $n \equiv \phi_{\mathrm{total}}/(2\pi)$, where $\phi_{\mathrm{total}}$ is the total change in azimuthal angle of the photon orbit \cite{Gralla:2019xty}. In detail, for an observer at the north pole, photons are categorized into three classes: (\romannumeral1) direct emission ($n < 3/4$): the photons cross the equatorial plane once; (\romannumeral2) lensing ring emission ($3/4 \leq n < 5/4$): the photons cross the equatorial plane twice; (\romannumeral3) photon ring emission ($n \geq 5/4$): the photons cross the accretion disk at least three times, as shown in the top row of Fig.~\ref{fig:nb+trajectories}. The distribution of light rays around the Schwarzschild, RN and non-minimally coupled black holes has been shown in the bottom row of Fig.~\ref{fig:nb+trajectories}, where the solid black disk stands for the black hole and the black dashed circle denotes the photon sphere, while the black, gold, and red curves correspond to direct emission, lensing ring emission, and photon ring emission, respectively. The exact ranges of $b$ for the direct, lensing ring, and photon ring emissions are also listed in Table~\ref{tb:b_range}.
\begin{table}[htbp]
	\centering
	\begin{tabular}{|c|c|c|c|c|}
		\hline
		{$Q/M$} & {$\xi/M^4$} & {Direct: $n < 3/4$} & {Lensing ring: $3/4 < n < 5/4$} & {Photon ring: $n > 5/4$} \\
		\hline
		{$0$} & {$0$} & {$b/M \notin (5.01514, 6.16757)$} & {$b/M \in (5.01514, 5.18781) \text{ or } (5.22794, 6.16757)$} & {$b/M \in (5.18781, 5.22794)$} \\
		\hline
		{$0.5$} & {$0$} & {$b/M \notin (4.77294, 5.97448)$} & {$b/M \in (4.77294, 4.95793) \text{ or } (5.00390, 5.97448)$} & {$b/M \in (4.95793, 5.00390)$} \\
		\hline
		{$0.5$} & {$0.5$} & {$b/M \notin (4.55564, 5.95238)$} & {$b/M \in (4.55564, 4.85255) \text{ or } (4.92534, 5.95238)$} & {$b/M \in (4.85255, 4.92534)$} \\
		\hline
	\end{tabular}
	\caption{The ranges of impact parameter $b$ corresponding to the direct emission, lensing ring emission, and photon ring emission for the Schwarzschild $(Q = 0,\ \xi = 0)$, RN $(Q = 0.5,\ \xi = 0)$, and non-minimally coupled $(Q = 0.5,\ \xi = 0.5)$ black holes.}
    \label{tb:b_range}
\end{table}

The results in Fig.~\ref{fig:nb+trajectories} and Table~\ref{tb:b_range} indicate that comparing with the Schwarzschild and RN black holes, the non-minimal coupling has an impact on the widths of various emission components. It is of particular interest to further investigate how these effects and the accretion position manifest in the observed intensities and, thus, in the overall optical appearance of the black hole surrounded by a thin accretion disk. Moreover, the above studies also show that the parameters $Q$ and $\xi$ have similar effects on both the motions of accretion flow and photons, therefore, we expect no degeneracy in the accretion images among the non-minimally coupled black hole and Schwarzschild black hole. Consequently, in what follows, we focus primarily on the differences in images of inclined accretion disks around the Schwarzschild, RN, and non-minimally coupled black holes, where the latter two share the same rescaled charge.

\section{Images of the black holes illuminated by an inclined thin accretion disk}\label{sec:disk_images}

In this section, we consider an inclined thin accretion disk extending to the event horizon of the black hole as the light source. In this case, the accreted massive particles either move in stable circular orbits with radii $r \ge r_{\mathrm{ISCO}}$ or move along the plunging orbits in the region of $r_{+} < r < r_{\mathrm{ISCO}}$.

\subsection{Emission model}\label{subsec:emission_model}

\begin{figure}[htbp]
	\centering
	\includegraphics[height=6.5cm]{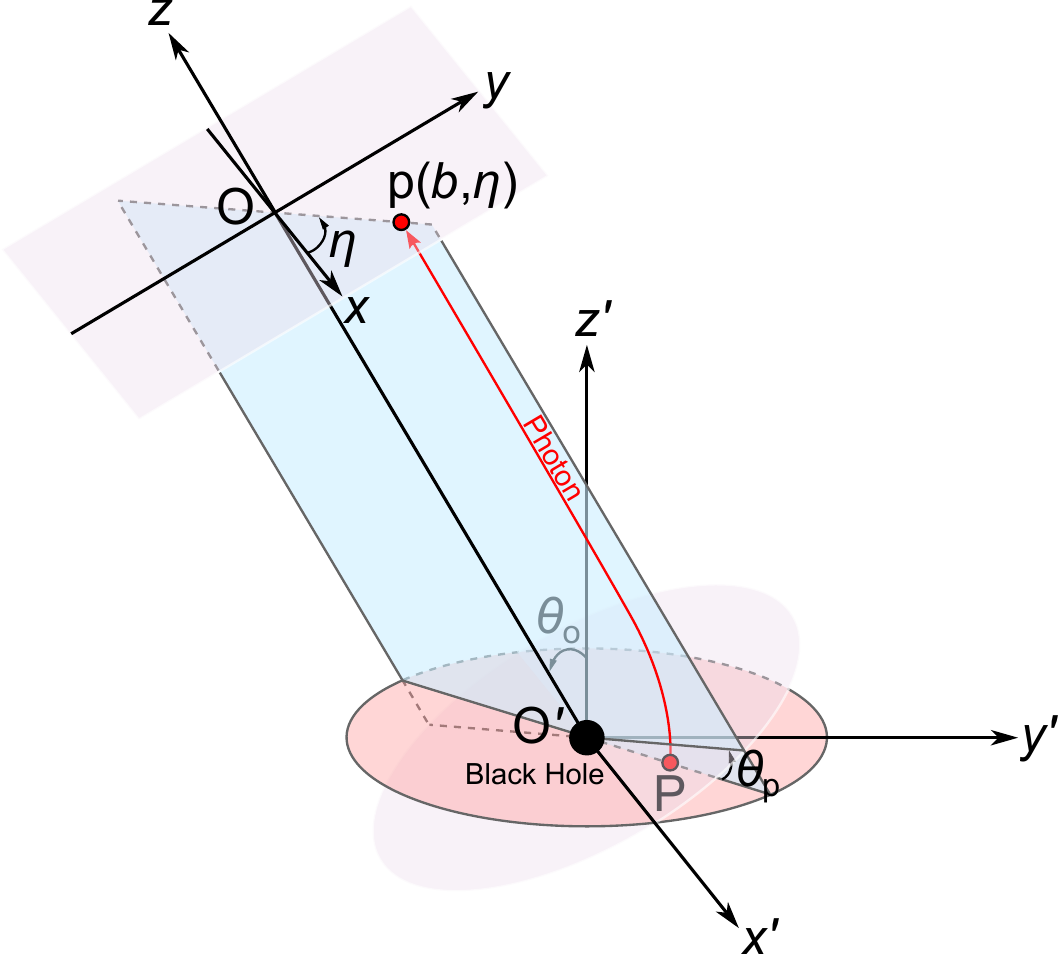}\hspace{1cm}
	\includegraphics[height=6.5cm]{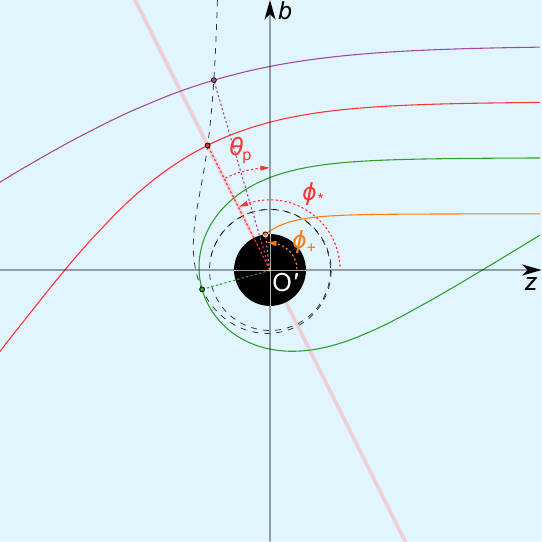}
	\caption{\textbf{Left panel:} the schematic diagram of the coordinate systems used in the current study, comprises the observation coordinate system $\mathrm{O}\text{-}xyz$ and the black hole coordinate system $\mathrm{O'}\text{-}x'y'z'$. The red plane and blue plane represent the equatorial plane and the photon trajectory plane, respectively. The two light violet transparent planes correspond to the observation plane and a parallel one centered on the black hole. The observation plane is inclined at an angle $\theta_{\mathrm{o}}$ relative to the equatorial plane. The photon trajectory plane intersects the equatorial plane and the observation plane along two lines, the angle between which is denoted by $\theta_{\mathrm{p}}$. A photon emitted from the source point $\mathrm{P}$ on the equatorial plane travels along the red trajectory and reaches the image point $\mathrm{p}$ on the observation plane, which is located at the polar coordinates $(b,\eta)$. \textbf{Right panel:} the photon trajectory plane with four light rays plotted by the orange, green, red, and violet solid curves. The central disk represents the black hole. The light red line denotes the equatorial plane where the accretion disk lies. The black dashed curve shows the periastron positions for orbits with different $b$, and the dot on each light ray marks its periastron.}
	\label{fig:coord_sys+periastron}
\end{figure}

We use a phenomenological model with the equatorial approximation given by Ref.~\cite{Gralla:2020srx}, assuming that all emissions come from the equatorial plane, to avoid the complex simulation of geometrically thick flows without losing utility in practice. The coordinate systems we use are shown in the left panel of Fig.~\ref{fig:coord_sys+periastron}, which depicts a red equatorial plane, a blue photon trajectory plane, and a light violet transparent observation plane accompanied by a parallel one centered on the black hole. The observed intensity at the polar coordinates $(b,\eta)$ in the observation plane inclined at an angle $\theta_{\mathrm{o}}$ relative to the equatorial plane, takes the form \cite{Gralla:2020srx,Chael:2021rjo},
\begin{equation}
\label{eq:Iobs}
	I_{\mathrm{obs}}(b, \eta, \theta_{\mathrm{o}}) = \sum_{m=1}^{N_{\mathrm{max}}} f_{m} g^3(r_{m}, b, \eta, \theta_{\mathrm{o}}) I_{\mathrm{em}}(r_{m}),
\end{equation}
where $r_{m}$ denotes the radius of the $m^{\text{th}}$ intersection between a light ray and the accretion disk, $N_{\mathrm{max}}$ is the maximum time of the intersecting, $g(r_{m}, b, \eta, \theta_{\mathrm{o}})$ is the redshift factor, $I_{\mathrm{em}}(r_{m})$ is the emission intensity at the radius of $r_{m}$, and $f_{m}$ is a ``fudge factor" to adjust the brightness of the higher-order rings.

Following the model given by Ref.~\cite{Chael:2021rjo}, which fits the 230 GHz time-averaged GRMHD simulation result, we set the fudge factor to
\begin{equation}
\label{eq:fm}
	f_{m} = 
	\begin{cases}
		1, &\quad m = 1, \\
		\frac{2}{3}, &\quad m > 1,
	\end{cases}
\end{equation}
and adopt the emission intensity profile
\begin{equation}
\label{eq:Iem}
	I_{\mathrm{em}}(r) = 
	\begin{cases}
		0, &\quad r < r_{+}, \\
		e^{-2\ln{\frac{r}{r_{+}}} - \frac{1}{2} \left( \ln{\frac{r}{r_{+}}} \right)^2}, &\quad r \ge r_{+}.
	\end{cases}
\end{equation}
The emission intensity peaks at the event horizon, decreases rapidly with radius, and drops to a very low level beyond $r_{\mathrm{ISCO}}$. Additionally, a smaller event horizon radius $r_{+}$ leads to a faster decay of the emission intensity.

\subsection{Transfer function}\label{subsec:transfer_function}

Before exploring the redshift in Eq.~\eqref{eq:Iobs}, we have to analyze $r_{m}$, which is also referred to as the transfer function, as it characterizes the mapping from the photon's impact parameter to its $m^{\text{th}}$ hitting position on the disk \cite{Gralla:2019xty}. For face-on images, $r_{m}$ is confirmed by seeking the point at which the azimuthal angle $\phi$ of the backward-traced photon trajectory satisfies
\begin{equation}
\label{eq:rm_condition_0}
    \frac{\phi}{2\pi} = \frac{2m-1}{4}.
\end{equation}
In the more general case with a non-zero inclination angle $\theta_{\mathrm{o}} \ne 0^{\circ}$ between the equatorial plane and the observation plane, the condition \eqref{eq:rm_condition_0} should be modified as
\begin{equation}
\label{eq:rm_condition_i}
    \frac{\phi - \theta_{\mathrm{p}}}{2\pi} = \frac{2m-1}{4},
\end{equation}
with
\begin{equation}
\label{eq:theta_p}
	\theta_{\mathrm{p}} = \arctan (\tan \theta_{\mathrm{o}} \sin \eta).
\end{equation}
Here, $\eta$ is the polar angle of the image point on the observation plane, and $\theta_{\mathrm{p}}$ denotes the angle between the intersection lines of the photon trajectory plane with the equatorial plane and the observation plane (see Fig.~\ref{fig:coord_sys+periastron}).

\begin{figure}[htbp]
	\centering
	\subfigure[\, $r_{1}$]
	{\includegraphics[height=4.8cm]{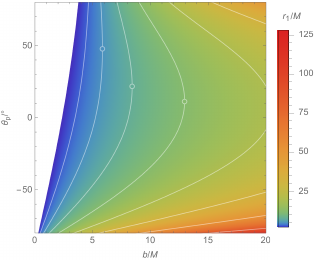}}\hspace{0.2cm}
	\subfigure[\, $r_{2}$]
	{\includegraphics[height=4.8cm]{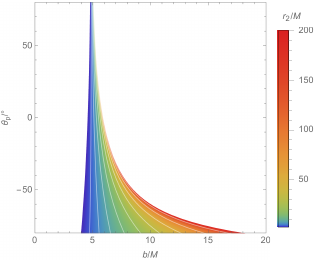}}\hspace{0.2cm}
	\subfigure[\, $r_{3}$]
	{\includegraphics[height=4.8cm]{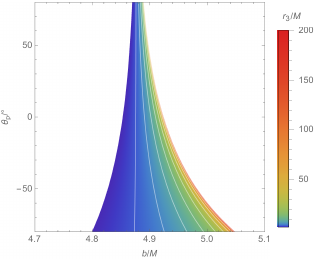}}
	\caption{The first three transfer functions $r_{1}(b, \theta_{\mathrm{p}})$, $r_{2}(b, \theta_{\mathrm{p}})$, and $r_{3}(b, \theta_{\mathrm{p}})$ for the non-minimally coupled black hole with $Q = 0.5,\ \xi = 0.5$. Here, the settings of $\theta_{\mathrm{p}}$ ranging from $-80^\circ$ to $80^\circ$, and the upper limit of $r_{m}$ to $200M$ are chosen to balance completeness and computational efficiency. The white contour lines in each plot indicate the values of $r_{+} \mathrm{e}^{n/2}$ for $n \in \mathbb{Z}^+$. The open circles in (a) mark the vertices of the contour lines, each of which corresponds to the periastron of a photon orbit.}
	\label{fig:transfunc_2D}
\end{figure}

Therefore, the transfer function $r_{m}$ is a function of both the impact parameter $b \in [0, \infty)$, which determines the shape of the photon orbit, and the angle $\theta_{\mathrm{p}} \in [-90^\circ, 90^\circ]$, which controls the inclination between the photon orbit and the accretion disk. The first three transfer functions for the non-minimally coupled black hole with $Q = 0.5,\ \xi = 0.5$ are plotted in Fig.~\ref{fig:transfunc_2D}. Here, we take $\theta_{\mathrm{p}}$ that ranges from $-80^\circ$ to $80^\circ$, and set the upper limit of $r_{m}$ to $200M$, balancing completeness and computational efficiency. Note that to explicitly show the tendency, we employ a rainbow color scheme based on an exponential distribution and highlight the contour lines with $r_m = r_{+} \mathrm{e}^{n/2}$ for $n \in \mathbb{Z}^+$ in each plot. It is obvious that, an increase in $\theta_{\mathrm{p}}$ leads to a reduction in the range of $b$, indicating that the width of each emission becomes narrower.

In fact, recalling the relation \eqref{eq:rm_condition_i}, we can stitch those plots of $r_{1}$, $r_{2}$, $r_{3}$ (even for larger $m$) together into a continuous whole by joining the edge $\theta_{\mathrm{p}} = 90^\circ$ of one to the edge $\theta_{\mathrm{p}} = -90^\circ$ of the next, although the regions of $\lvert \theta_{\mathrm{p}} \rvert > 80^\circ$ are not shown in Fig.~\ref{fig:transfunc_2D}. Subsequently, we treat the transfer functions of all $m$ collectively as a single entity, denoted by $r_{m}$. According to Fig.~\ref{fig:transfunc_2D}, we see that for a fixed $\theta_{\mathrm{p}}$, $r_{m}$ increases with $b$. However, the variation of $r_{m}$ with $\theta_{\mathrm{p}}$ for fixed $b$ is a bit subtle, which motivates us to present those quantities in the photon trajectory plane shown in the right panel of Fig.~\ref{fig:coord_sys+periastron}.

For a given $b$ with $b > b_{\mathrm{c}}$, the value of $r_{m}$ reaches its minimum at the vertex of the related contour line, as marked by the open circle in Fig.~\ref{fig:transfunc_2D}(a). This minimum is naturally equal to the radius of the periastron for the corresponding photon orbit. Thus, at the vertex, we have $\theta_{\mathrm{p}} = \phi_{*} - (m-1/2)\pi$, where $\phi_{*}$ is the azimuth of the periastron. This is described by the red light ray in the right panel of Fig.~\ref{fig:coord_sys+periastron}. Similarly, for light rays with different $b$ (e.g., the violet and green ones), we can find their corresponding $\theta_{\mathrm{p}}$ and $\phi_{*}$ at the vertex. Therefore, we can conclude that for a fixed $b > b_{\mathrm{c}}$, $r_{m}$ always increases as $\theta_{\mathrm{p}}$ deviates from $\phi_{*} - (m-1/2)\pi$. Note that the value of $\phi_{*}$ tends to $\pi/2$ with $b \to \infty$, and tends to infinity rapidly with $b \to b_{\mathrm{c}}$. 

For a given $b$ with $b < b_{\mathrm{c}}$, the minimum of $r_{m}$ is nothing but the event horizon radius $r_{+}$, because this photon orbit always terminates at the event horizon. Thus, in this case, the maximums of $\theta_{\mathrm{p}}$ we can read from Fig.~\ref{fig:transfunc_2D} is the intersect between the vertical line for fixed $b$ and the left edge (i.e., the contour line for $r_{m} = r_{+}$). According to the right panel of Fig.~\ref{fig:coord_sys+periastron}, this maximum of $\theta_{\mathrm{p}}$ is equal to $\phi_{+} - (m-1/2)\pi$. Here $\phi_{+}$ is the azimuth where the photon orbit with the given $b$ terminates at the event horizon, see the orange light ray. Therefore, for a fixed $b < b_{\mathrm{c}}$, when $\theta_{\mathrm{p}}$ exceeds its maximum, the corresponding light ray does not contribute to the image because it has no intersection with the accretion disk, otherwise, it shall extract the energy from the disk at $r_{m}$ which increases monotonically as $\theta_{\mathrm{p}}$ decreases from its maximum $\phi_{+} - (m-1/2)\pi$. Obviously, we have $\phi_{+} = 0$ when $b = 0$, and $\phi_{+}$ tends to infinity rapidly with $b \to b_{\mathrm{c}}$.

\begin{figure}[htbp]
	\centering
	\subfigure[\, $\theta_{\mathrm{p}} = 0^\circ\ \text{with}\ \theta_{\mathrm{o}} = 0^{\circ}$]
	{\includegraphics[height=3.5cm]{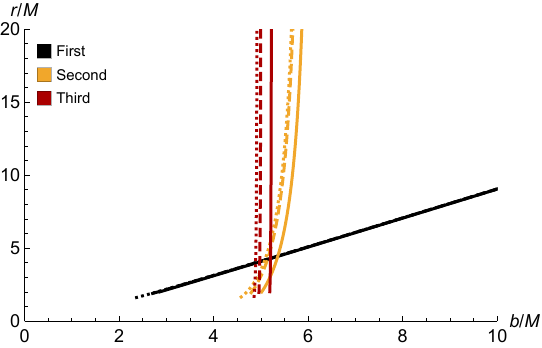}}\hspace{0.5cm}
	\subfigure[\, $\theta_{\mathrm{p}} = 80^\circ\ \text{with}\ \theta_{\mathrm{o}} = 80^{\circ}, \eta = \frac{1}{2}\pi$]
	{\includegraphics[height=3.5cm]{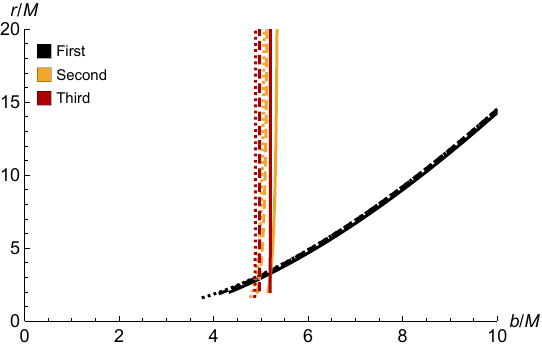}}\hspace{0.5cm}
	\subfigure[\, $\theta_{\mathrm{p}} = -80^\circ\ \text{with}\ \theta_{\mathrm{o}} = 80^{\circ}, \eta = \frac{3}{2}\pi$]
	{\includegraphics[height=3.5cm]{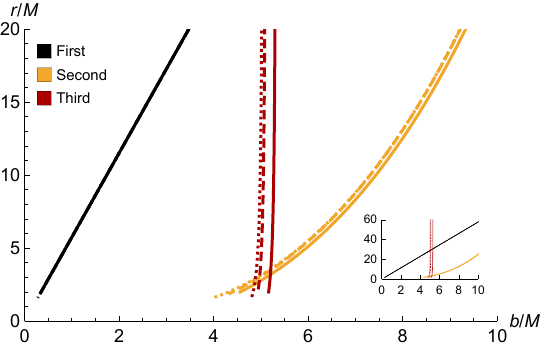}}
	\caption{The transfer functions with different sets of inclination angle $\theta_{\mathrm{o}}$ and polar angle $\eta$ (i.e., different $\theta_{\mathrm{p}}$). The solid, dashed, and dotted lines respectively correspond to the Schwarzschild black hole, RN black hole with $Q = 0.5$, and non-minimally coupled black hole with $Q = 0.5,\ \xi = 0.5$, while the black, gold, and red colors denote the first, second, and third intersections with the disk, respectively.}
    \label{fig:transfunc_b}
\end{figure}

To see the effects of black hole parameters, we depict the first three transfer functions with three different $\theta_{\mathrm{p}}$ for the non-minimally coupled black hole against the Schwarzschild and RN black holes in Fig.~\ref{fig:transfunc_b}. It is shown that the transfer functions for the non-minimally coupled black hole and RN black hole are shifted to the left by certain distances relative to that of the Schwarzschild black hole, with the shift of the RN black hole smaller than that of the non-minimally coupled black hole. Such shifts result in corresponding reductions in image size. When $\theta_{\mathrm{o}} = 0^{\circ}$, the orbital inclination $\theta_{\mathrm{p}}$ as well as the transfer function shown in Fig.~\ref{fig:transfunc_b}(a), are already determined due to the rotational invariance with respect to $\eta$. When the inclination angle increases to $\theta_{\mathrm{o}} = 80^{\circ}$, as shown in Fig.~\ref{fig:transfunc_b}(b) and Fig.~\ref{fig:transfunc_b}(c), the slopes of the red and gold lines decrease for $\eta = \pi/2$ and increase for $\eta = 3\pi/2$ compared with the case $\theta_{\mathrm{o}} = 0^{\circ}$. The slope $\partial r_{m}/\partial b$ quantifies the demagnification factor at each $b$, thus these opposite trends in the slopes explain the differential deformation between the upper and lower parts of both the photon ring and lensing ring emissions. 

\begin{figure}[htbp]
	\centering
	\includegraphics[width=17.5cm]{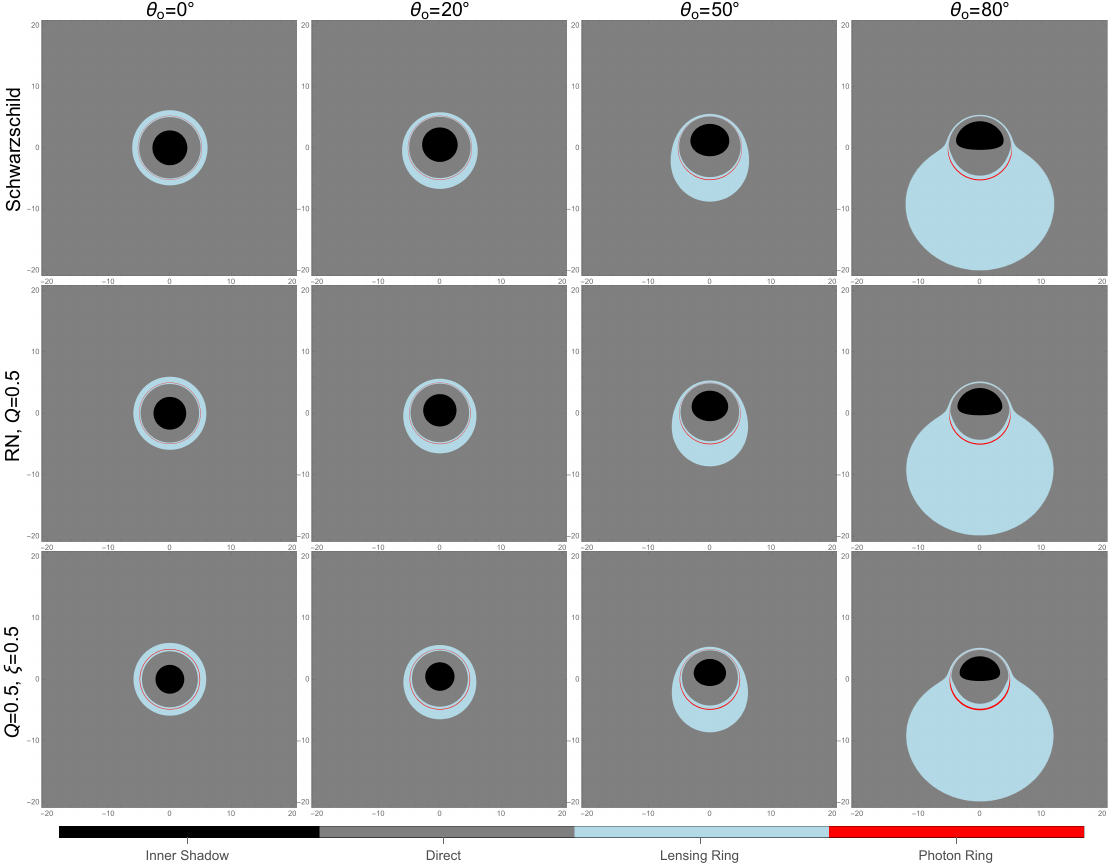}
	\caption{The image regions of accretion disks surround the Schwarzschild, RN, and non-minimally coupled black holes at varying inclination angles, characterized by different emission types. The black, gray, cyan, and red regions correspond to the inner shadow, direct emission, lensing ring emission, and photon ring emission, respectively.}
    \label{fig:regions}
\end{figure}

The range of $b$ for the $m^{\text{th}}$ transfer function is exponentially suppressed with $m$ \cite{Gralla:2019xty}. So we shall focus on the first three orders of emissions, since the higher cases contribute even much less to the total luminosity. The spatial distribution of the first three orders of emissions in the images is shown in Fig.~\ref{fig:regions}, where the black, gray, cyan, and red colors correspond to the inner shadows, direct emissions, lensing ring emissions, and photon ring emissions, respectively. The figure shows that the inner shadow, from which the backward-traced rays reach the event horizon without intersecting the accretion disk, becomes progressively smaller from the Schwarzschild black hole to the RN black hole to the non-minimally coupled black hole, while the widths of the lensing ring and photon ring increase in the same sequence. As $\theta_{\mathrm{o}}$ increases, the inner shadows flatten, and the upper parts of the images become narrower, as well as the lower parts become wider.

\subsection{Redshift factor}\label{subsec:redshift_factor}

\begin{figure}[bp]
	\centering
	\includegraphics[width=17.5cm]{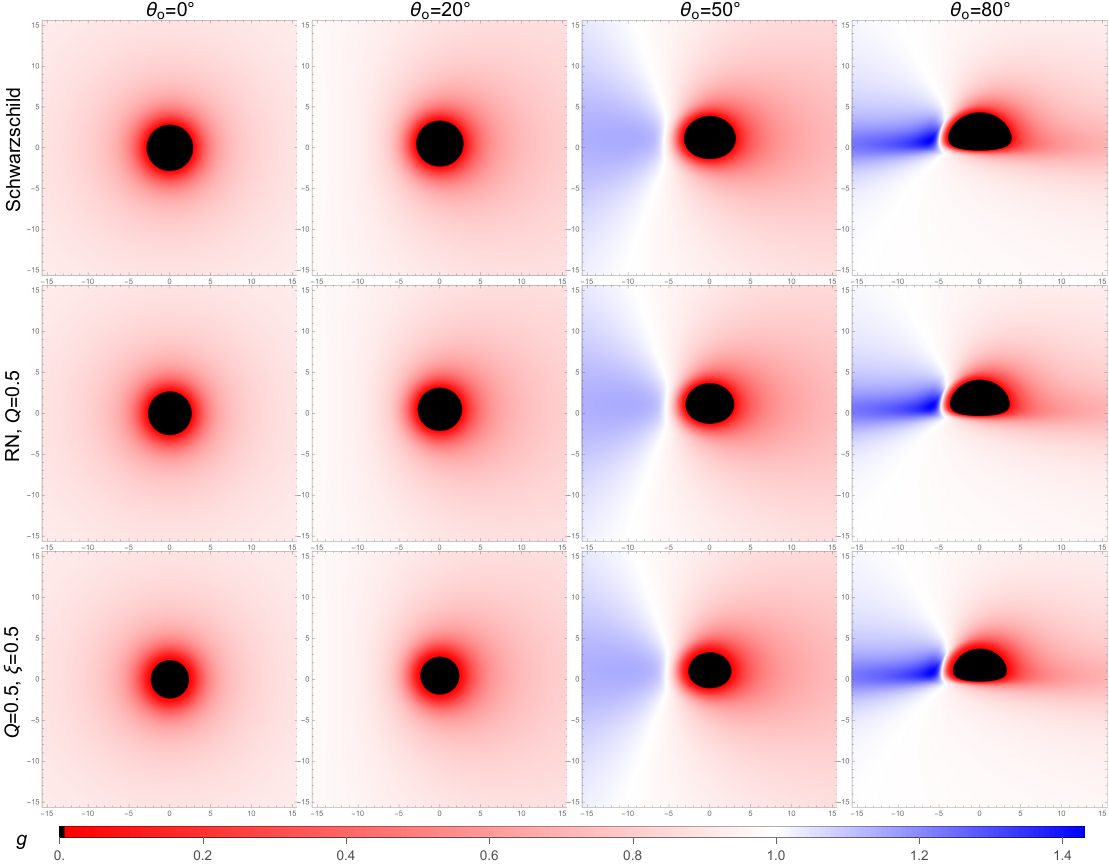}
	\caption{The redshift factor distributions for the direct emission from accretion disks surround the Schwarzschild, RN, and non-minimally coupled black holes at varying inclination angles.}
    \label{fig:redshift1}
\end{figure}
\begin{figure}[bp]
	\centering
	\includegraphics[width=17.5cm]{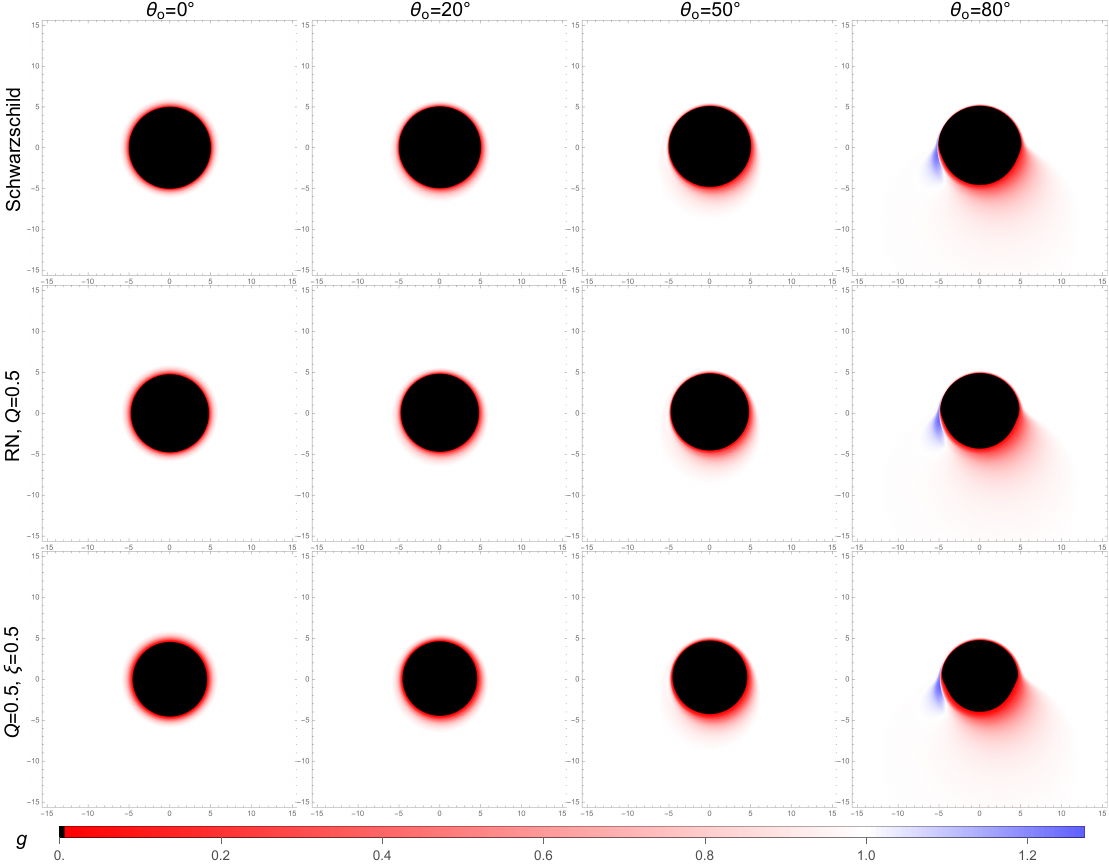}
	\caption{The redshift factor distributions for the lensing ring emission from accretion disks surround the Schwarzschild, RN, and non-minimally coupled black holes at varying inclination angles.}
    \label{fig:redshift2}
\end{figure}

To figure out the image, a critical aspect is to determine the redshift factor of the light. The generic expression for the redshift factor reads:
\begin{equation}
\label{eq:redshift_general}
	g = \frac{p_\mu u_{\mathrm{obs}}^\mu}{p_\nu u_{\mathrm{em}}^\nu},
\end{equation}
where $p_{\mu}$ is the conjugate momentum of the photon, $u_{\mathrm{obs}}^\mu = (1, 0, 0, 0)$ is the four-velocity of the static distant observer, and $u_{\mathrm{em}}^\nu = (\dot{t}_{\mathrm{em}}, \dot{r}_{\mathrm{em}}, \dot{\theta}_{\mathrm{em}}, \dot{\phi}_{\mathrm{em}})$ is the four-velocity of the emitting source.

As we have confined the accretion disk to the equatorial plane, thus $u_{\mathrm{em}}^\theta = 0$, and Eq.~\eqref{eq:redshift_general} can be simplified into
\begin{equation}
\label{eq:redshift_simplified}
	g = \left( u_{\mathrm{em}}^t + u_{\mathrm{em}}^r \frac{p_r}{p_t} + u_{\mathrm{em}}^\phi \frac{p_\phi}{p_t} \right)^{-1}.
\end{equation}
For the general case shown in the left panel of Fig.~\ref{fig:coord_sys+periastron}, using the definition \eqref{eq:p_mu}, we have
\begin{equation}
	\frac{p_\phi}{p_t} = b \sin\theta_{\mathrm{o}} \cos\eta,
\end{equation}
following Ref.~\cite{Luminet:1979nyg}. Then, based on the null condition $g^{\mu\nu} p_{\mu} p_{\nu} = 0$ and the spherical symmetry of spacetime, we obtain the following equation
\begin{equation}
	\frac{p_r}{p_t} = -\sqrt{-\frac{g^{tt}}{g^{rr}} - \frac{g^{\phi \phi}}{g^{rr}} b^2}.
\end{equation}

In addition to $p_{\mu}$, we have to calculate the redshift factor \eqref{eq:redshift_simplified} determined by the $t$, $r$, and $\phi$ components of the four-velocity of the emitting source as it orbits in different paths. In the simplified model we used, the accreted massive particles either move in stable circular orbits with radii $r \ge r_{\mathrm{ISCO}}$ or move along the plunging orbits in the region of $r_{+} < r < r_{\mathrm{ISCO}}$.

As discussed in Sec.~\ref{subsec:massive_particle}, for stable circular orbits with radii $r \ge r_{\mathrm{ISCO}}$, the $t$, $r$, and $\phi$ components of the four-velocity of the emitting source are
\begin{equation}
	u_{\mathrm{em}}^{t} = -\frac{E}{g_{tt}}, \quad u_{\mathrm{em}}^{\phi} = \frac{L}{g_{\phi\phi}}, \quad u_{\mathrm{em}}^{r} = 0,
\end{equation}
where $E$ and $L$ are given by Eq.~\eqref{eq:SCO_E-L-O}. For plunging orbits in the region of $r_{+} < r < r_{\mathrm{ISCO}}$, with conserved energy $E = E_{\mathrm{ISCO}}$ and angular momentum $L = L_{\mathrm{ISCO}}$, we obtain
\begin{equation}
	u_{\mathrm{em}}^{t} = -\frac{E_{\mathrm{ISCO}}}{g_{tt}}, \quad u_{\mathrm{em}}^{\phi} = \frac{L_{\mathrm{ISCO}}}{g_{\phi\phi}}, \quad u_{\mathrm{em}}^{r} = -\sqrt{\frac{1}{g_{rr}} \left( -1 - \frac{E_{\mathrm{ISCO}}^2}{g_{tt}} - \frac{L_{\mathrm{ISCO}}^2}{g_{\phi\phi}} \right)},
\end{equation}
where $u_{\mathrm{em}}^{r}$ is solved from the time-like normalization condition $g_{\mu \nu} u_{\mathrm{em}}^{\mu} u_{\mathrm{em}}^{\nu} = -1$, and the negative sign indicates that the particle is plunging into the black hole.

Now we are prepared to calculate the redshift factor \eqref{eq:redshift_simplified}. The contributions of the redshift factors for the first two orders of emissions from accretion disks surround the Schwarzschild, RN, and non-minimally coupled black holes are shown in Figs.~\ref{fig:redshift1} and \ref{fig:redshift2}. The black regions of Fig.~\ref{fig:redshift1} correspond to the inner shadows, while in Fig.~\ref{fig:redshift2}, the black regions encompass both the inner shadows and a portion of direct emissions surrounded by lensing ring emissions, as illustrated by the central black and gray areas in Fig.~\ref{fig:regions}. For this region, the backward-traced rays intersect the accretion disk at most once and finally reach the event horizon. The redshift characteristics of the photon ring emissions are nearly invisible, so we also plot the redshift factors for all three emissions along the $x$-axis and $y$-axis in Figs.~\ref{fig:redshift1_sec}, \ref{fig:redshift2_sec}, and \ref{fig:redshift3_sec} which are exhibited in the Appendix~\ref{appendix}. As $\theta_{\mathrm{o}}$ increases, the motion of accretion disk creates a considerable blueshifted region for the direct emission, and smaller regions for the lensing ring emission and photon ring emission.

\begin{figure}[htbp]
	\centering
	\includegraphics[width=6.5cm]{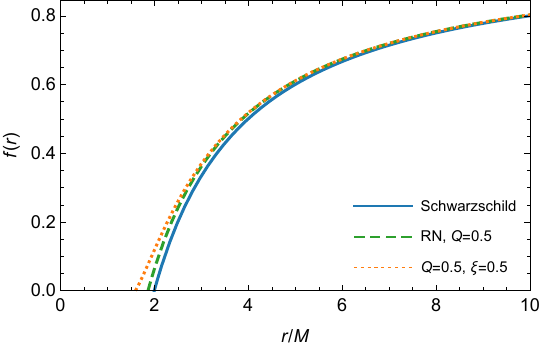}\hspace{1.5cm}
    \includegraphics[width=6.5cm]{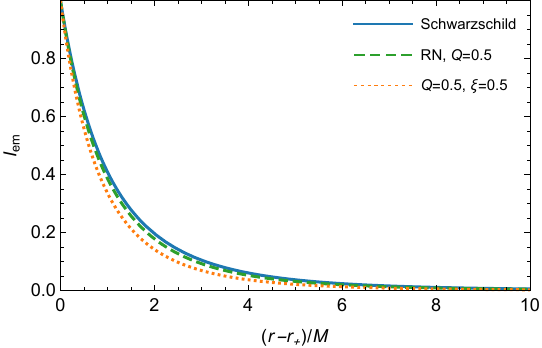}
	\caption{\textbf{Left panel:} the profile of the metric function $f(r)$ with $r \ge r_{+}$ for the Schwarzschild, RN, and non-minimally coupled black holes. \textbf{Right panel:} the distributions of the emission intensity for the Schwarzschild, RN, and non-minimally coupled black holes.}
    \label{fig:fr+IemCompare}
\end{figure}

Compared with the other two types of black holes, the presence of the non-minimal coupling parameter enhances the redshift effects to some extent. As shown in Fig.~\ref{fig:redshift1}, the highly redshifted regions near the inner shadow are significantly wider for the non-minimally coupled black hole than for the RN and Schwarzschild black holes. Such differences can also be found for the lensing ring and photon ring emissions, and are more clearly manifested in Figs.~\ref{fig:redshift1_sec}--\ref{fig:redshift3_sec} that, in the inner regions, the redshift factors for the non-minimally coupled black hole decrease slower than those for the RN and Schwarzschild black holes. We argue that this phenomenon is directly caused by the behavior of the metric function $f(r)$, based on the following facts: (\romannumeral1) it appears in the inner regions; (\romannumeral2) it is independent of the inclinations; (\romannumeral3) our previous studies indicate that the effects of $\xi$ on $E_{\mathrm{ISCO}}$, $L_{\mathrm{ISCO}}$ and $r_{m}$ show no specificity compared to those of $Q$. To illustrate this, we plot the profile of $f(r)$ with $r \ge r_{+}$ for the Schwarzschild, RN, and non-minimally coupled black holes in the left panel of Fig.~\ref{fig:fr+IemCompare}. The plot reveals that near the event horizon, the slope of $f(r)$ for the non-minimally coupled case is significantly smaller than that for the other two, which is consistent with the previously noted slower decrease in the redshift factors. We also checked the case of a static disk, where the redshift factor is simply $\sqrt{f(r)}$, and the results are qualitatively the same.

\subsection{Optical appearance}\label{subsec:optical_appearance}

\begin{figure}[htbp]
	\centering
	\includegraphics[width=17.5cm]{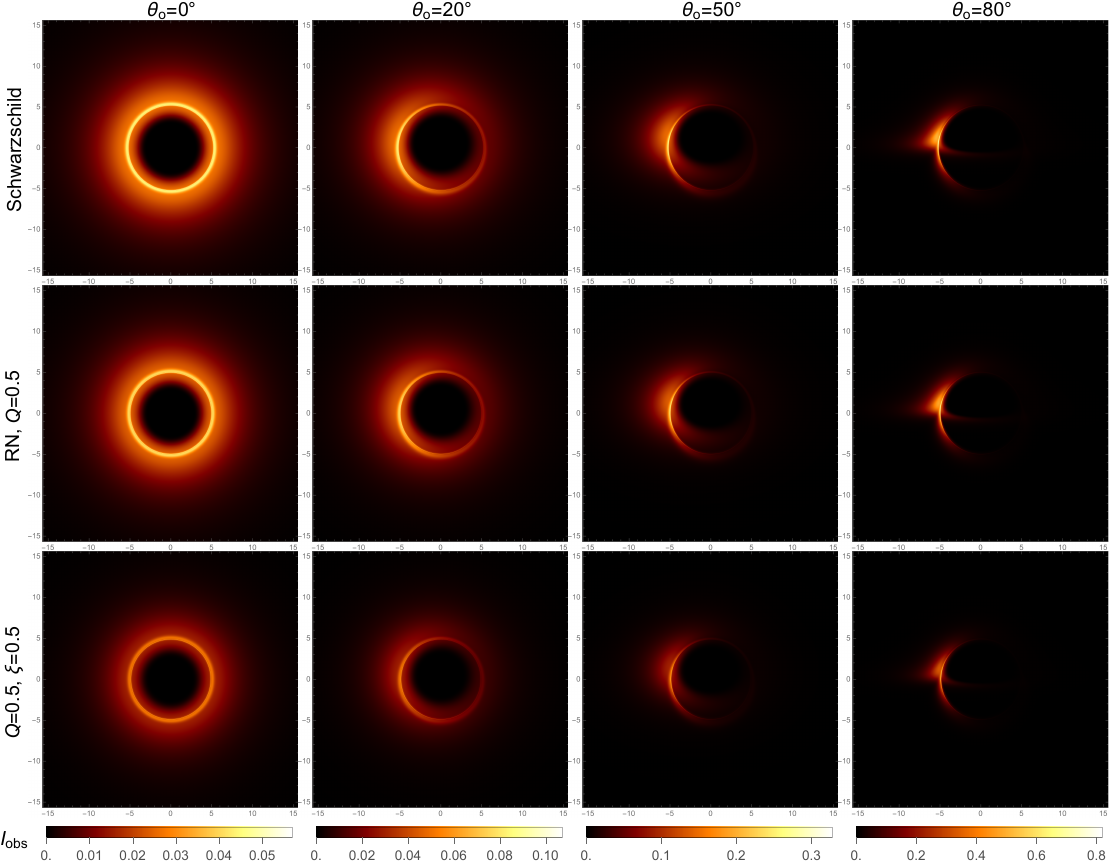}
	\caption{The optical appearances of the Schwarzschild, RN, and non-minimally coupled black holes at varying inclination angles.}
    \label{fig:images_SOPLS}
\end{figure}

The optical appearances of the Schwarzschild black hole, RN black hole with $Q = 0.5$, and non-minimally coupled black hole with $Q = 0.5,\ \xi = 0.5$, are plotted in Fig.~\ref{fig:images_SOPLS}. We select four inclination angles, which are $0^{\circ}$, $20^{\circ}$, $50^{\circ}$, and $80^{\circ}$. The image is shown to be a symmetric ring when the inclination is $0^{\circ}$, with a bright photon ring at the radius of $b_{\mathrm{c}}$. The dark parts of the center, i.e., the inner shadows, become non-circular when the inclination is $20^{\circ}$, as well as the higher brightness in the left part than in the right because of the Doppler shift caused by the motion of the accretion disk. With the inclination turning to $50^{\circ}$, the direct emission of the disk appears as a bright band crossing the shadow, and the right part becomes much darker than the left part. When the angle of inclination changes to a high value of $80^{\circ}$, the difference in brightness makes the right part almost invisible. While the angle of inclination increases, the position of the photon ring remains virtually unchanged, as well as the upper part becomes thinner, and the lower part becomes thicker, which have been shown more clearly in Fig.~\ref{fig:regions}.

\begin{figure}[htbp]
	\centering
	\includegraphics[width=4.3cm]{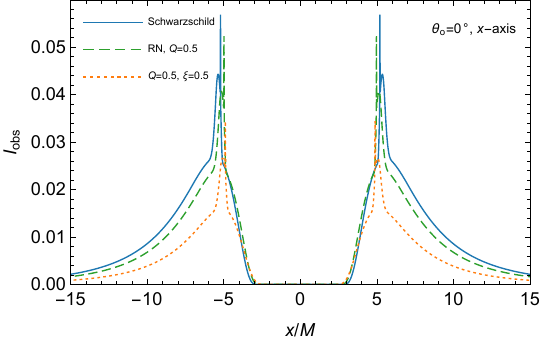}
    \includegraphics[width=4.3cm]{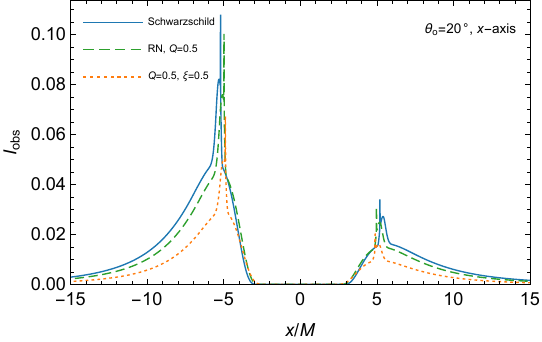}
    \includegraphics[width=4.3cm]{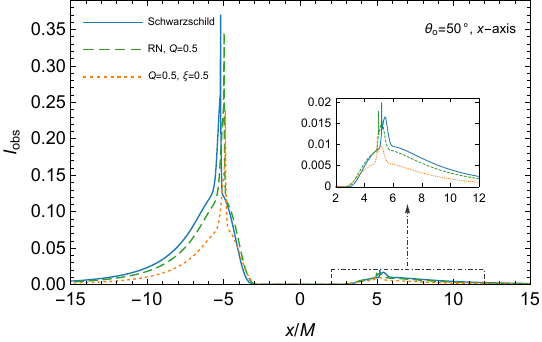}\hspace{0.0895cm}
    \includegraphics[width=4.2105cm]{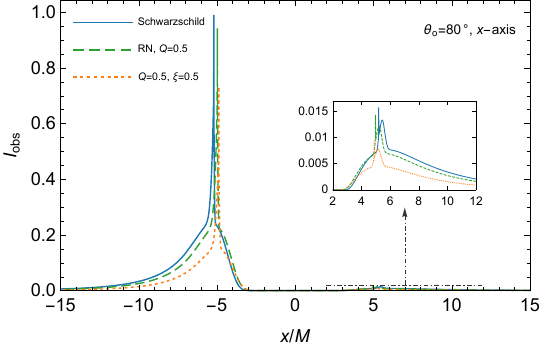}\\
	\includegraphics[width=4.3cm]{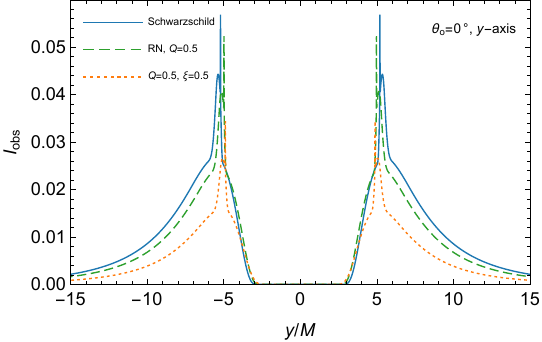}
	\includegraphics[width=4.3cm]{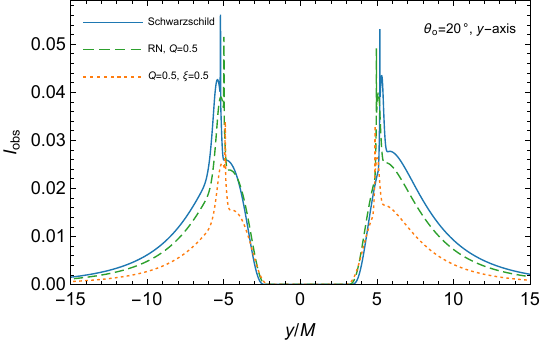} 
	\includegraphics[width=4.3cm]{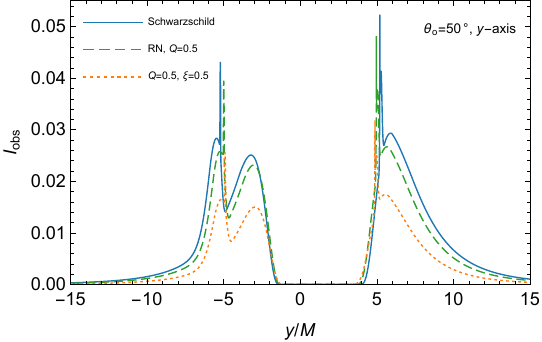} 
	\includegraphics[width=4.3cm]{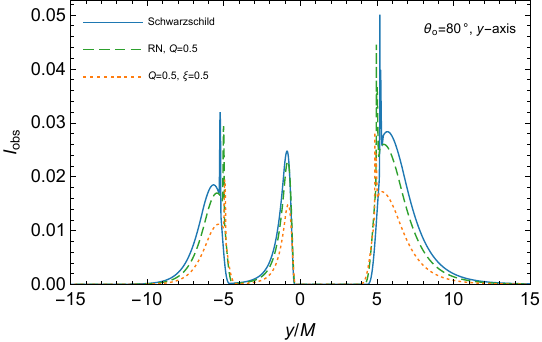}
	\caption{The observed intensities along the $x$-axis ($\eta = 0\ \text{and}\ \pi$) and $y$-axis ($\eta = \pi/2\ \text{and}\ 3\pi/2$) for the Schwarzschild, RN, and non-minimally coupled black holes at varying inclination angles.}
    \label{fig:compare_sec}
\end{figure}
The differences between the images of the Schwarzschild, RN, and non-minimally coupled black holes are manifested mainly in the total size and brightness, that the images of the non-minimally coupled black hole are smaller and significantly darker than those of the RN black hole with the same $Q$, while the images of the Schwarzschild black hole are the largest and brightest. To clearly quantify the difference in size and brightness, we plot Fig.~\ref{fig:compare_sec}, which shows the observed intensities along the $x$-axis ($\eta = 0\ \text{and}\ \pi$) and $y$-axis ($\eta = \pi/2\ \text{and}\ 3\pi/2$) of the Schwarzschild, RN, and non-minimally coupled black holes with inclination angles $\theta_{\mathrm{o}} = 0^{\circ},\ 20^{\circ},\ 50^{\circ},\ \text{and}\ 80^{\circ}$. The figure reveals that the non-minimal coupling makes the images significantly darker and further reduces their sizes. The differences in the total brightness are caused by the dependence of the emission intensity on $r_{+}$ (see Eq.\eqref{eq:Iem}), which is determined by the black hole parameters $Q$ and $\xi$ (see Fig.~\ref{fig:rph+bc+reh}(c)). Specifically, the behavior of the emission intensity in the right panel of Fig.~\ref{fig:fr+IemCompare} obviously shows that the existence of $Q$ and $\xi$ increases its decay rate. Furthermore, it can be observed that as the inclination angle increases, a third peak emerges in the $y$-axis profile, which is consistent with the information presented in Fig.~\ref{fig:images_SOPLS}.

\section{Conclusions}\label{sec:conclusion}

In this paper, we have investigated the imaging characteristics of an exact spherically symmetric static regular black holes with a thin accretion disk within the framework of non-minimally coupled $SU(2)$ Einstein-Yang-Mills theory with a Wu-Yang ansatz. The disk was considered to consist of free, electrically neutral plasma, wherein matter follows stable circular orbits outside the ISCO, while inside the ISCO, it rapidly plunges into the black hole. We first analyzed how the coupling constants influence the orbital dynamics of both massive and massless particles in this spacetime background. Based on this analysis, we systematically compared the redshift factors of the accretion disk and the resulting black hole images in the spacetimes of the Schwarzschild, RN, and non-minimally coupled black holes. Our results demonstrate that the presence of the non-minimal coupling parameter has imprints on the observable features of the black hole image.

More specifically, to investigate the imaging characteristics of the non-minimally coupled black hole with a thin accretion disk, it is essential to first analyze the dynamical behaviors of both massive and massless particles orbiting the black hole. This involves examining the influence of the non-minimal coupling constant $\xi$ on geodesic motion starting from the metric functions $f(r)$ of the black hole. Our results show that, for both massive and massless particles, the effect of $\xi$ on the stable circular orbits is nearly linear: as $\xi$ increases, the orbital radius, along with the corresponding energy and angular momentum for massive particles or the impact parameter for photons, decreases approximately linearly, see Figs.~\ref{fig:ISCO} and \ref{fig:rph+bc+reh}. In contrast, the event horizon radius $r_{+}$ exhibits a nonlinear monotonic decrease with increasing coupling constant. In particular, as $\xi$ approaches its critical value, the event horizon radius decreases sharply with increasing $\xi$. Furthermore, comparing with the Schwarzschild and RN black holes, the non-minimal coupling has an impact on the widths of various emission components.

Based on this analysis, we modeled an inclined thin accretion disk extending down to the event horizon as the light source. Using an emission model that better aligns with the observational data and numerical simulations, we calculated the redshift distribution of photons emitted by the disk, and compared the results for the non-minimally coupled black hole with those for the Schwarzschild and RN black holes. Compared with the latter two cases, the region of high redshift near the inner shadow in the non-minimally coupled black hole spacetime is wider, indicating that the presence of the non-minimal coupling parameter can enhance the gravitational redshift effect to some extent. Furthermore, the observed intensities from the non-minimally coupled black hole are lower than those for the other two cases due to the strong impact of the coupling parameter on the event horizon, as shown in Fig.~\ref{fig:images_SOPLS}.

Although these results elucidate the imaging characteristics of the non-minimally coupled black hole in the presence of a thin accretion disk, providing a theoretical foundation for further understanding and potential observational tests of non-minimal EYM theory, we emphasize that this work is based on certain idealized assumptions. For instance, the accretion disk model considered here significantly simplifies the complexity of realistic astrophysical accretion disks. Furthermore, our findings require validation through future observations with higher precision, such as the Black Hole Explorer mission \cite{Johnson:2024ttr,Lupsasca:2024xhq}. However, this study still provides a potential way to distinguish the non-minimal coupling black hole from static black holes in GR. We also expect that these preliminary results could shed light on the future test of non-minimal EYM theory using other observational messengers.

\section*{Acknowledgments} 
This work is partly supported by the Natural Science Foundation of China under Grant No. 12375054. Y.-Z. Li is also supported by the research funds No. 2055072312.

\appendix
\section{Redshift factors along the x-axis and y-axis}\label{appendix}

\begin{figure}[htbp]
	\centering
	\includegraphics[width=4.3cm]{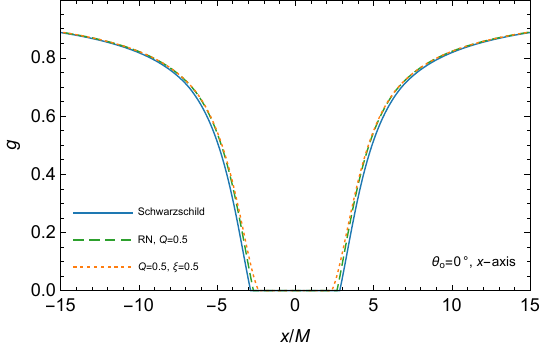}
    \includegraphics[width=4.3cm]{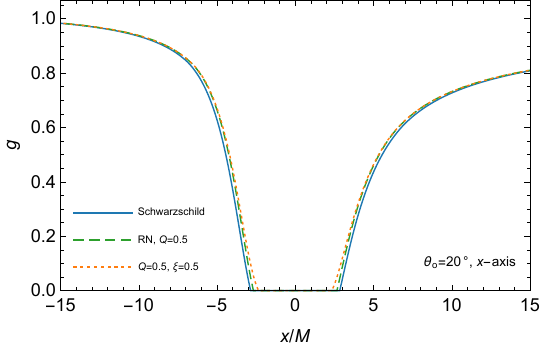}
    \includegraphics[width=4.3cm]{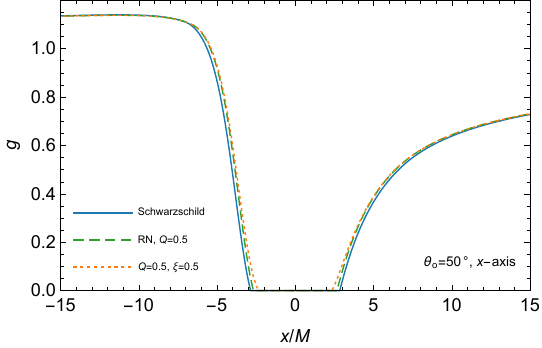}
    \includegraphics[width=4.3cm]{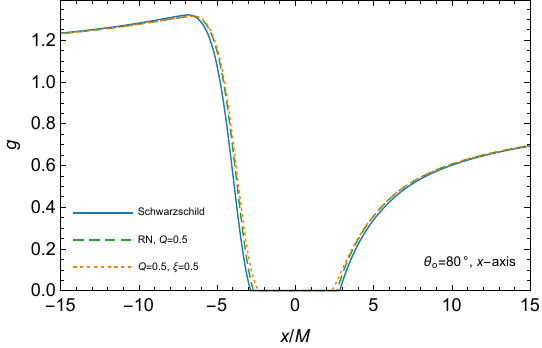} \\
	\includegraphics[width=4.3cm]{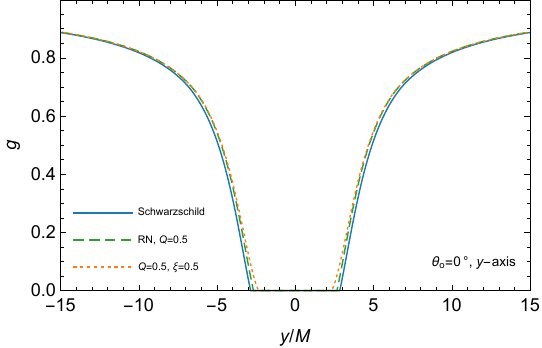}
	\includegraphics[width=4.3cm]{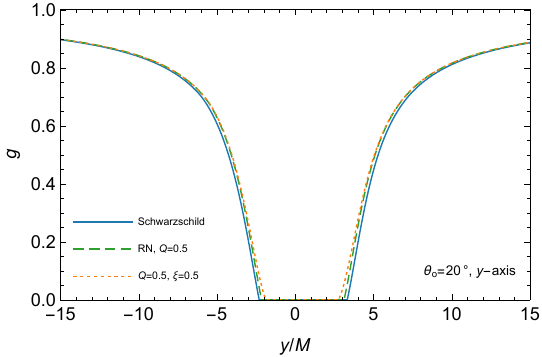} 
	\includegraphics[width=4.3cm]{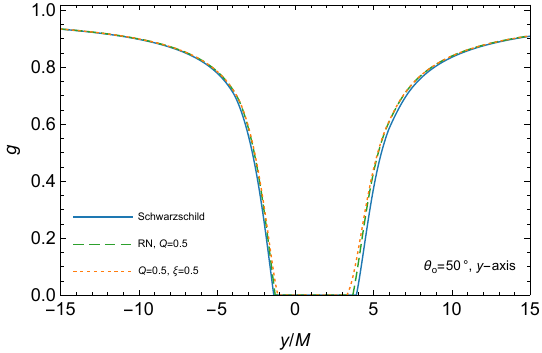} 
	\includegraphics[width=4.3cm]{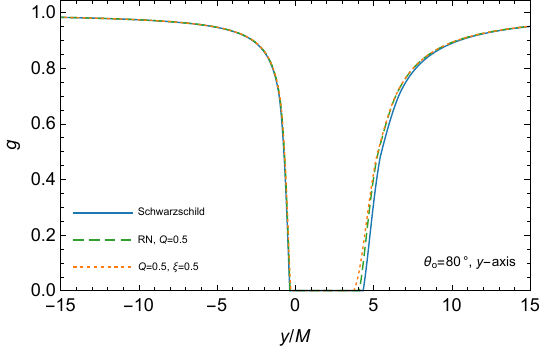}
	\caption{The redshift factors of the direct emission along the $x$-axis ($\eta = 0\ \text{and}\ \pi$) and $y$-axis ($\eta = \pi/2\ \text{and}\ 3\pi/2$) for the Schwarzschild, RN, and non-minimally coupled black holes at varying inclination angles.}
    \label{fig:redshift1_sec}
\end{figure}

\begin{figure}[htbp]
	\centering
	\includegraphics[width=4.3cm]{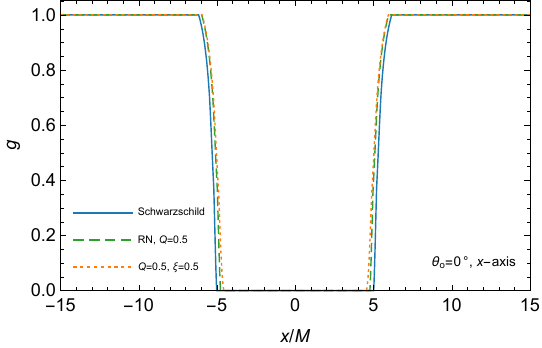}
    \includegraphics[width=4.3cm]{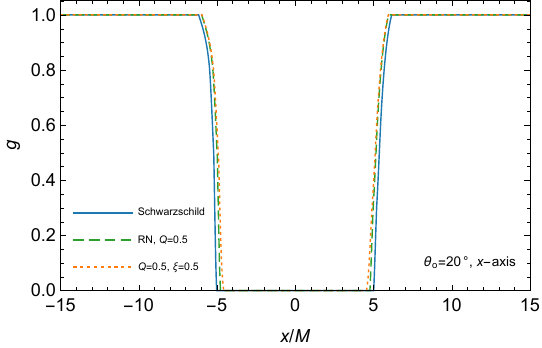}
    \includegraphics[width=4.3cm]{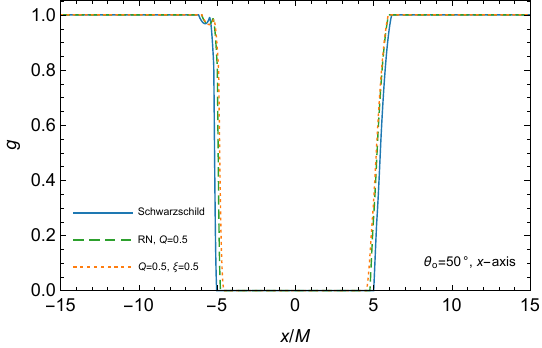}
    \includegraphics[width=4.3cm]{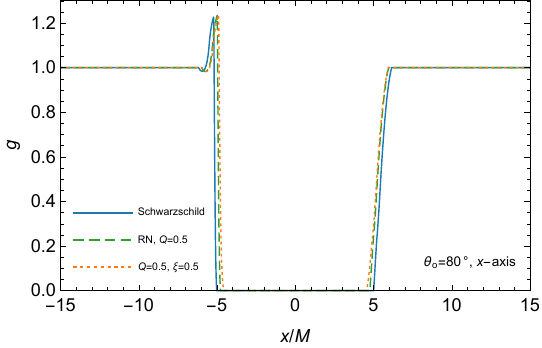}\\
	\includegraphics[width=4.3cm]{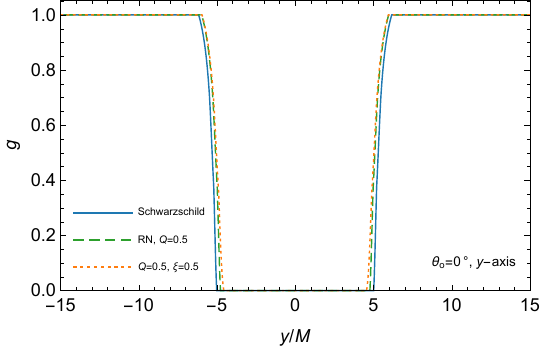} 
	\includegraphics[width=4.3cm]{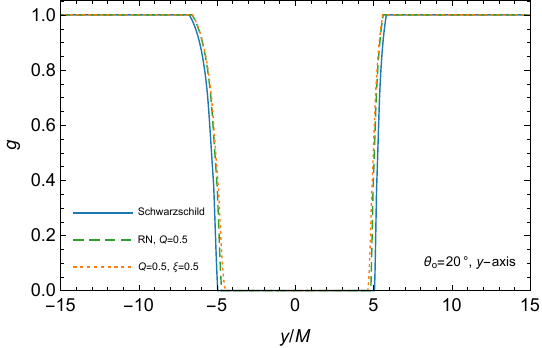}
	\includegraphics[width=4.3cm]{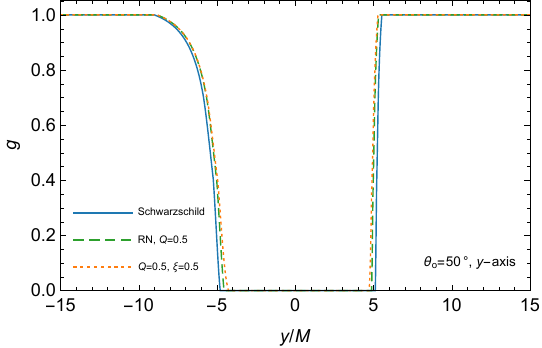} 
	\includegraphics[width=4.3cm]{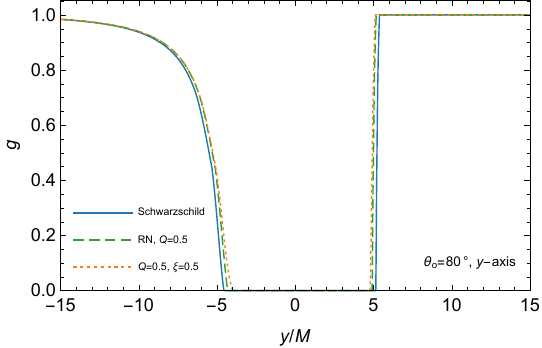}
	\caption{The redshift factors of the lensing ring emission along the $x$-axis ($\eta = 0\ \text{and}\ \pi$) and $y$-axis ($\eta = \pi/2\ \text{and}\ 3\pi/2$) for the Schwarzschild, RN, and non-minimally coupled black holes at varying inclination angles.}
    \label{fig:redshift2_sec}
\end{figure}

\begin{figure}[htbp]
	\centering
	\includegraphics[width=4.3cm]{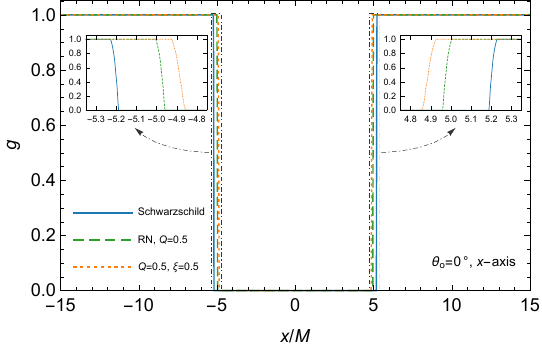}
    \includegraphics[width=4.3cm]{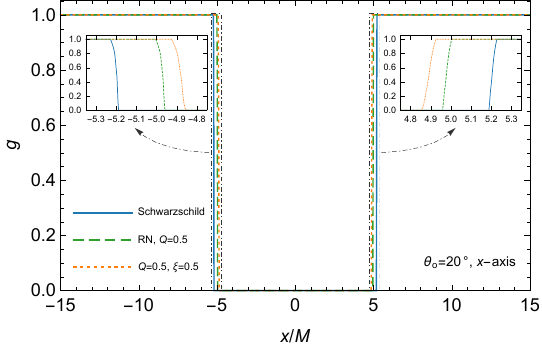}
    \includegraphics[width=4.3cm]{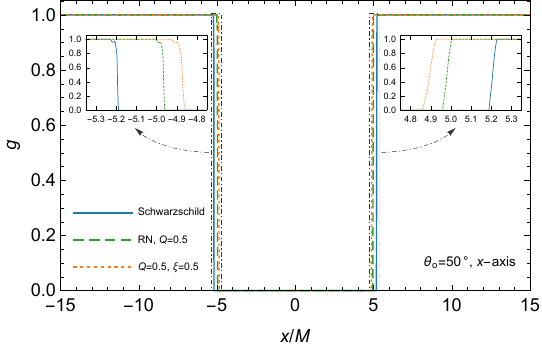}
    \includegraphics[width=4.3cm]{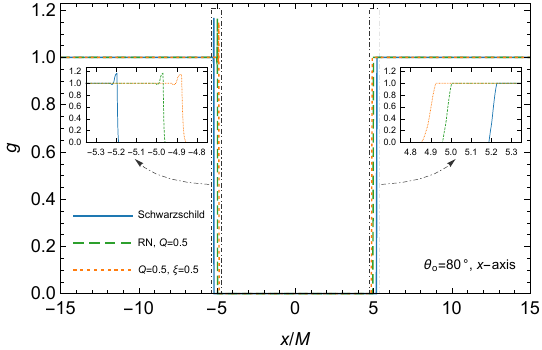}\\
	\includegraphics[width=4.3cm]{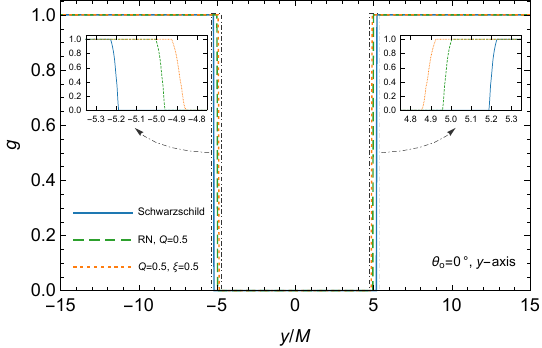} 
	\includegraphics[width=4.3cm]{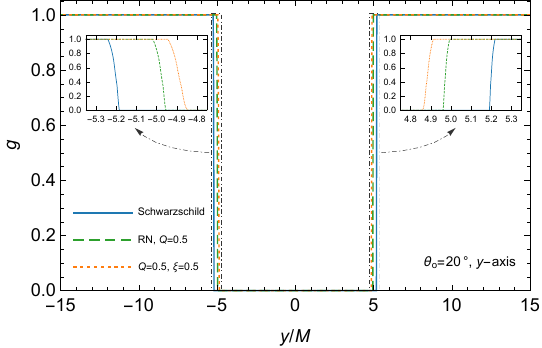} 
	\includegraphics[width=4.3cm]{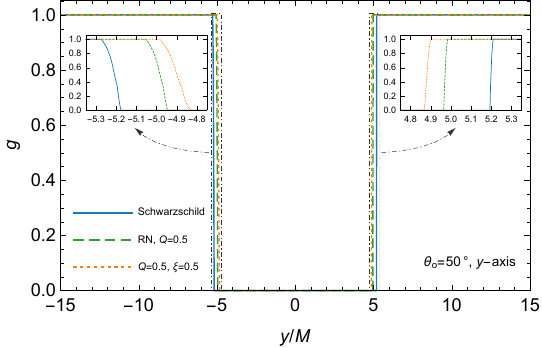}
	\includegraphics[width=4.3cm]{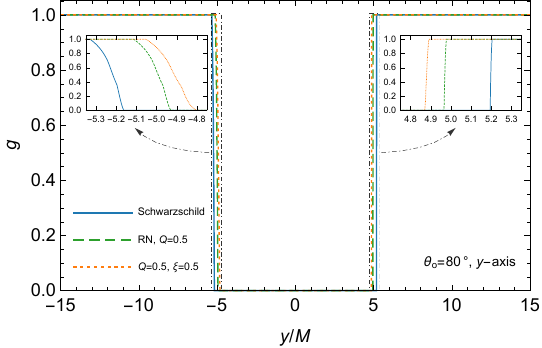}
	\caption{The redshift factors of the photon ring emission along the $x$-axis ($\eta = 0\ \text{and}\ \pi$) and $y$-axis ($\eta = \pi/2\ \text{and}\ 3\pi/2$) for the Schwarzschild, RN, and non-minimally coupled black holes at varying inclination angles.}
    \label{fig:redshift3_sec}
\end{figure}

In this appendix, we present the redshift factors for all three emissions along the $x$-axis and $y$-axis in Figs.~\ref{fig:redshift1_sec}, \ref{fig:redshift2_sec}, and \ref{fig:redshift3_sec}.

\bibliographystyle{utphys}
\bibliography{ref}

\end{document}